\documentclass[12pt]{JHEP3}

\usepackage{amsmath,epsfig}
\usepackage{amssymb,amsfonts}
\usepackage{latexsym}
\usepackage{longtable}
\relax
\renewcommand{\theequation}{\arabic{section}.\arabic{equation}}
\def\be{\begin{equation}}
\def\ee{\end{equation}}
\def\bea{\begin{eqnarray}}
\def\eea{\end{eqnarray}}

\newcommand\fverb{\setbox\pippobox=\hbox\bgroup\verb}
\newcommand\fverbdo{\egroup\medskip\noindent%
                        \fbox{\unhbox\pippobox}\ }
\newcommand\fverbit{\egroup\item[\fbox{\unhbox\pippobox}]}

\newcommand{\bear}{\begin{eqnarray}}

\newcommand{\eear}{\end{eqnarray}}

\newbox\pippobox

\def\6{\partial}

\def\a{\alpha}
\def\nn{\nonumber}

\def\pa{\partial}

\def\e{\epsilon}
\def\m{\mu}
\def\n{\nu}

\def\s{\sigma}

\def\sp{\;\;\;,\;\;\;}

\def\sq
\def\a{\alpha}

\def\l{\lambda}

\def\hri#1#2{\href{http://arxiv.org/abs/#1}{[ArXiv:#1]#2}}
\def\hre#1#2{\href{http://arxiv.org/abs/#1/#2}{[ArXiv:#1/#2]}}

\def\na{\nabla}
\def\aa{|A|}
\def\cc{|C|}

\def\e{\epsilon}

\title{On Hor\v ava-Lifshitz ``Black Holes"}
\author{{\large Elias Kiritsis\footnote{On leave
of absence from APC, Universit\'e Paris 7, (UMR du CNRS 7164)}, Georgios Kofinas }
~\\
~\\
Crete Center for Theoretical Physics,
Department of Physics,\\ University of Crete,
71003 Heraklion, Greece}

\preprint{ CCTP-2009-16}

\abstract{The most general spherically symmetric solution with zero shift is found
in the non-projectable Ho\v rava-Lifshitz class of theories
with general coupling constants.
It contains as special cases, spherically symmetric solutions found by other authors earlier.
It is found that the generic solution has conventional (AdS, dS or flat) asymptotics with a universal $1/r$ tail.
There are several special cases where the asymptotics differ, including the detailed balance choice of couplings. 
The conventional thermodynamics of this general class of solutions is
 established by calculating the energy, temperature and entropy.
Although several of the solutions have conventional horizons, for
 particles with ultra-luminal dispersion relations such solutions appear to be horizonless. }

\begin{document}

\section{Introduction and results}
\label{intro}
The UV completion of gravity has been a difficult road for theoretical physics of the past fifty years.
The only convincing answer so far has been provided by string theory, and it works only
in perturbation theory and at energies well below the Planck scale.

Recently, a different field theory model for a UV complete theory of
gravity was proposed \cite{hor1} (see also \cite{klu}).
The theory  does not have the full diffeomorphism
invariance of GR but only a subset. Because of this property a different scaling
 symmetry is allowed in the UV and the theory accepts renormalizable
couplings with up to six derivatives.
It is interesting that this picture allows a theory of gravitation that is scale-invariant in the UV, and where standard
general relativity with its higher symmetry
could be an emerging theory in the IR.

There are several versions by now of the Ho\v rava-Lifshitz type of theory.
Two main categories were introduced originally by Ho\v rava. The non-projectable version allows the lapse $N$
to be a general function of spacetime coordinates. In the projectable version, $N$ is a function of time only.

Furthermore in the original formulation the principle of detailed balance (DB) has been imposed. It had the advantage of
reducing the possible coupling constants of the theory.
In subsequent works \cite{kk,svw}, it was advocated that a general action should be used that allows all couplings compatible
with renormalizability.
It was first argued in \cite{nastase} that DB implies a correlation between the effective Planck scale and cosmological constant that left
the theory very little room to agree with observation.

It was subsequently  shown in \cite{pope}, that the large distance gravitational field of a spherically symmetric source has a very different behavior in the HL
theory with detailed balance than GR, a fact that excludes HL gravity with detailed balance as a description of low energy observable gravity.
At the same time,  a small breaking of the detailed balance reinstated the usual $1/r$ tail of gravity \cite{pope}.

It is obvious that we have a family of HL-type theories with projectable or non-projectable lapse, and with or without detailed balance.
In view of this, we will reiterate the philosophy advocated in \cite{kk}: the most general HL theory compatible with renormalizability
should be investigated, not only in terms of its consistency but also in terms of its ability to agree with observables in gravity and cosmology.

The arena where the theory seems most promising is cosmology. It was initially pointed out that the UV structure of the theory can help in solving the
horizon and flatness problems without inflation  \cite{kk}. In particular, the fact that speed of light in the UV is infinite indicates there is no
horizon problem, while the fact that curvature contributions to the Friedmann equation are enhanced, indicates that they must be
 suppressed at early times. This is similar to standard holographic cosmology \cite{holo}, although the mechanism for the generation
 of the curvature square corrections is different.

 Scale invariant perturbations can be generated naturally in the HL high-energy phase without
inflation \cite{muko1}, \cite{kk}. Relevant terms provide power suppressed corrections to the scale invariant spectrum. It was argued in \cite{kk}
that the logarithmic running of marginal couplings in the UV will provide a
 logarithmic tilt in the scale invariant spectrum of perturbations, although
the direction of the tilt cannot be determined without a detailed calculation.

The presence of higher derivative terms in the theory provide the opportunity of a bouncing universe \cite{Cal}, \cite{kk}, \cite{bra}.
The existence of parity odd couplings at high energy provide the possibility of polarization asymmetry in the CMB data \cite{ts}.
It should be noted that CP non-invariance is not a necessary condition of the formalism, but it is certainly allowed.
It is present in the original Ho\v rava proposal.

In the projectable theory, the Hamiltonian constraint is not a local constraint. It was shown in \cite{muko3} that in the cosmological context
the constraint allows dark matter as an integration constant in the FRW equations.
Many further analyses of cosmological issues have been done since \cite{Piao}.

The issue of spherically symmetric solutions and black holes, as well as issues related to thermodynamics,
 were  analyzed in several works \cite{pope}, \cite{ks}-\cite{bh}.
Other issues involved strong coupling problems \cite{Cai}-\cite{blas2},
 comparison with solar system data \cite{test1},
issues of quantum field theory and renormalizability \cite{qft1}, as well as the discussion of particle geodesics \cite{poly}-\cite{rama}.

An important issue is related to the reduced diffeomorphism invariance and the associated strong coupling problems.
As anticipated in \cite{kk}, the dynamics of the extra scalar mode that exists in the theory is particular and can lead to strong coupling problems.
This was argued to be the case first in \cite{charmousis} and most convincingly in \cite{blas}. Moreover, generic perturbative instabilities
were found in the renormalizable regime of couplings \cite{bogdanos}. The projectable theory seems to be free of such problems, but
is expected to have problems with caustics in the cosmological domain \cite{blas}. In \cite{muko4} arguments against caustics formation were advanced.
Recently a modified theory, where additions of spatial derivatives of the lapse  were utilized, was argued to be problem-free \cite{blas2}.

An important issue for this class of gravity theories remains open. This is the issue of quantum UV structure and renormalizability.
Although the theory is power-counting renormalizable, several of the important UV couplings are dimensionless. They will therefore have
logarithmic UV running, and renormalizability requires that such couplings are asymptotically free.
Although this can always be guaranteed by taking the appropriate sign for a given coupling,
positivity of such couplings may not allow the appropriate choice. For example, $\lambda>1$ for the phenomenological viability of the theory.
The $\beta$ functions for the UV marginal couplings have not yet been calculated and therefore this important issue remains unresolved.
In particular, the question whether $\l=1$ is an IR fixed point remains open.

The potential asymptotic freedom of the UV couplings of the theory may have an additional impact on the cosmological constant problem.
Indeed, it allows the possibility that an exponentially small vacuum energy is generated because of asymptotic freedom.

An important coupling in this theory is the speed of light. This is a relevant
coupling in the UV theory and therefore its RG running will be powerlike.
It is however a marginal coupling in the IR theory and therefore its IR asymptotics will have a logarithmic running.
This implies that the RG behavior cannot be followed from the UV to the IR by perturbation theory.

The breaking of Lorentz invariance in the gravitational theory leads to several important and in principle observable effects.
As was pointed-out first in \cite{kk} and subsequently investigated in many papers, the matter sector of the gravitational theory
will have Lorentz invariance broken, either by the UV couplings or by quantum effects
 that communicate Lorentz invariance breaking from the gravitational sector. There are
  severe constraints on such Lorentz violating couplings that have been mostly derived in the last decade or so \cite{mat}-\cite{liberati}.
A particularly important question is to what extend the speeds of light relevant for different particles will be equal.
Inequality of such speeds of light in the IR is strongly constrained.
On the other hand, the RG structure of non-relativistic matter theories is very interesting
\cite{anselmi}, and may also accommodate novel effects for the SM
\cite{anselmi2}.

Issues due to the broken Lorentz invariance  arise also in the study of spherically symmetric
solutions. Indeed, in such cases the notion of the horizon may be particle dependent
if Lorentz invariance is broken. This is an issue that can be studied using the appropriate generalization of geodesic equations as in
\cite{poly}-\cite{rama}. This however implies that the notion of temperature is therefore particle-dependent, if particles have dispersive geodesics.
Worse, the notion of thermodynamics becomes fuzzy as the analogue of Hawking radiation depends on dispersion and ceases to be exactly thermal
\cite{corley}, \cite{parentani}. Moreover instabilities appear through the form of perpetuum mobile \cite{sibiryakov}.

It has been speculated by many, that non-relativistic gravitation theories of the
Ho\v rava-Lifshitz type may serve as duals of non-relativistic strongly coupled
large\,-$N$ quantum field theories.
We find such expectations to be remote and only for field theories that break translation invariance.
Standard holography as described by standard (super-gravity) has all the ingredients to describe
translationally invariant non-relativistic quantum field theories of the Lifshitz type.
The reason is that such QFTs are different from Lorentz invariant ones by the choice of couplings in the Lagrangian.
 Indeed, by choosing non-Lorentz invariant sources breaking the symmetry between space and time, we can generate in the UV the appropriate
 non-relativistic QFT. In the dual (supergravity) language the equations remain the same (fully diffeomorphism invariant),
  only the sources change.

Another hint in the same direction is   provided by the intuition
that it is the translation invariance of a boundary QFT that is promoted to diffeomorphism invariance in the bulk
leading to standard gravitational bulk theories. As the diffeomorphism invariance of Ho\v rava-Lifshitz gravities is smaller, this suggests
that such a gravitational theory maybe dual to a QFT with reduced translation invariance.
A better understanding of the holographic role of Ho\v rava-Lifshitz gravity remains to be found.

The purpose of this paper is to find and analyze spherically
symmetric, static solutions in Ho\v rava-Lifshitz gravity. The
appropriate ansatz for the metric degrees of freedom can be written
as
\be
ds^2=-N(r)^2~dt^2+{(dr\!+\!N^r(r) dt)^2\over
f(r)}+r^2d\Omega_k^2\,,
\ee
where $\Omega_k$ is a two sphere of unit
radius when $k=1$, a torus of unit volume when $k=0$ and a
pseudosphere of unit radius when $k=-1$. Unlike the fully
diffeomorphic invariant case, we cannot set the shift $N^r$ to zero
by a coordinate transformation.

The general action of the Ho\v rava-Lifshitz class of theories
is\footnote{This general action was advocated in \cite{kk} on the
basis of analyzing the phenomenology of Ho\v rava-Lifshitz gravity.
In \cite{kk}, $\beta_1=0$ as contributions to FRW equations were
studied. $\beta_1\not=0$ was introduced in \cite{svw} where a
different basis of marginal terms is used.}
\be
S=\int
dtd^3x\sqrt{g}N\!\left[\alpha (K_{ij}K^{ij}\!-\!\l K^2)\!+\!\gamma
{\cal E}^{ijk}R_{il}\na_j{R^l}_{k} \!+\!\zeta R_{ij}R^{ij}\!+\!\eta
R^2\!+\!\xi R\!+\!\sigma\right]+S_3\,, \label{i112}
\ee
where $S_3$
is the part of the action with six derivatives
\be
S_3=\int
dtd^3x\sqrt{g}N\left[ \beta C_{ij}C^{ij}+\beta_1 R\square
R+\beta_2R^3+\beta_3RR_{ij}R^{ij}+\beta_4
R_{ij}R^{ik}{R^{j}}_k\right] .
\ee
We will make the simplifying
assumption in this paper that $\beta_1=\beta_2=\beta_3=\beta_4=0$.
The reason is that apart from simplifying the equations, the
qualitative structure of the solutions we will find is expected to be similar.
Indeed, the presence of all curvature square terms takes into account curvature non-linearities.

To make the problem tractable we will also look for solutions that have $N^r=0$.
In this case we will find the most general solutions to the non-linear equations of the theory.

We find that:

$\bullet$ It is a generic feature of the solutions found that they have
regular large distance asymptotics that are asymptotically AdS/dS or
flat. Moreover, generically the next correction is compatible with a
standard Newton's law.

There are exceptions to this result (in  special cases). In
particular, the detailed balance action first written down by Hor\v
ava is one of these notable exceptions. As shown already first in
\cite{pope}, it does not reproduce the correct Newton law at large
distances. This is part of a class of special cases analyzed here,
occurring when
\be
B=\frac{1}{3\zeta\!+\!8\eta}
\Big(\!\frac{3\xi^{2}}{2(\zeta\!+\!3\eta)}\!-\!2\sigma\!\Big)=0 ~~~{\rm or}~~~
C=\frac{16(\zeta\!+\!3\eta)}{3\zeta\!+\!8\eta}<0,
\ee
although in the second case it is not the generic solution.

$\bullet$ For the general solution we calculate the temperature, mass and entropy,
 assuming Lorentz-invariant probe particles with standard dispersion relations and using the first law.
In one of the two general categories of parameters ($A\leq 0$ with $A$ defined in (\ref{tbun7}))
where the horizon distance is permitted to shrink to zero, the
entropy has a logarithmic divergence in this limit. In the other general category
($A>0$) the horizon distance is bounded from below.

$\bullet$ For special values of the curvature-squared couplings
$\zeta+3\eta=0$, when the horizon position  is allowed  to go to zero,
the entropy is regular in this limit.

$\bullet$ For the same values of the curvature-squared couplings $\zeta\!+\!3\eta=0$,
there exists a solution with logarithmically corrected large distance asymptotics.
Its ``mass parameter" defined in the naive way, depends logarithmically on radial distance and becomes smaller at larger distances.

$\bullet$ When the cosmological constant term in the action is absent, the
effective cosmological constant of (A)dS curvature is also zero
(for some branches).

$\bullet$ We study geodesics of particles with finite and infinite light speeds in spherically symmetric backgrounds
with traditional horizons. We parameterize the dispersion relations of particles as $p_0^2=(\vec p^{\,2})^n$, with $n\geq 1$.
We find that when $n>1$ the traditional behavior of the horizon disappears, suggesting that for such particles the black hole is
effectively ``naked".

The structure of this paper is as follows.

In section 2 we review the Ho\v rava-Lifshitz theory and its generalizations.

In section 3 we introduce the ansatz for the solutions, as well as the equations to be solved.
We also include a discussion of several important conceptual issues for such solutions ranging from the notion of horizon and singularity
to the notion of black-hole thermodynamics.

In  section 4 we discuss black hole solutions in the standard case, where higher curvature corrections are set to zero in order to
establish notation and connect to the standard cases.

In section 5 we discuss the spherically symmetric solutions in two special cases, determined by specific relations of the curvature-squared couplings.
In the first case our solutions are new. In the second, our solutions generalize previous solutions found by Lu, Mei and Pope \cite{pope},
Kehagias and Sfetsos \cite{ks}, and Park \cite{park}.

In section 6 we discuss the most general spherically symmetric solutions, by distinguishing two main cases involving the coupling constants.

In section 7 we describe the analysis of the conventional horizons for the general solutions.

In section 8 we calculate the conventional thermodynamic properties (for regular probes) assuming the validity of the first law.

Finally, section 9 contains a summary of the results as well as open questions.

In four appendices we collect several technical calculations.

Appendix A contains a derivation of the general equations of motion.

Appendix B fills-in the details for the derivation of the special class of solutions.

Appendix C fills-in the details for the derivation of the general class of solutions.

Finally, appendix D derives ``generalized geodesic equations" and solves for the radial geodesics.

\section{The Ho\v rava-Lifshitz gravity theory and generalizations}

We review here the Ho\v rava-Lifshitz gravitational theory as was formulated in \cite{hor1}
and as it was generalized in subsequent works \cite{kk,svw}.

The dynamical variables are $N,N_i,g_{ij}$, with scaling dimension
zero, except $N_i$ that has scaling dimension 2. This is similar to
the ADM decomposition of the metric in standard general relativity,
where the metric is written as
\be
ds^2=-N^2
~dt^2+g_{ij}(dx^i+N^idt)(dx^j+N^jdt)\sp N_i=g_{ij}N^j. \label{1}
\ee
The scaling transformation of the coordinates is now modified to
\be
t\to \ell^3~t\sp x^i\to \ell~x^i, \label{15}
\ee
under which $g_{ij}$
and $N$ are invariant, while $N^i$ scales as $N^i\to \ell^{-2} N_i$.

The kinetic terms are given by
\be S_K={2\over \kappa^2}\int dtd^3x
\sqrt{g}N\left(K_{ij}K^{ij}-\l K^2\right)\sp K=g^{ij}K_{ij}\sp
K^{ij}=g^{ik}g^{jl}K_{kl} \label{2}
\ee
in terms of the extrinsic curvature
\be
K_{ij}={1\over 2N}(\dot
g_{ij}-\nabla_{i}N_j-\nabla_jN_i), \label{3}
\ee
 with covariant
derivatives defined with respect to the spatial metric $g_{ij}$.

The most general power-counting renormalizable action is
\be
S=\int
dtd^3x\sqrt{g}N\!\!\left[\alpha (K_{ij}K^{ij}\!-\!\l K^2)\!+\!\gamma
{\cal E}^{ijk}R_{il}\na_j{R^l}_{k} \!+\!\zeta R_{ij}R^{ij}\!+\!\eta
R^2\!+\!\xi R\!+\!\sigma\!\right]+S_3, \label{i12}
\ee
with $ {\cal
E}^{ijk}={\e^{ijk}\over \sqrt{g}}$ the standard generally covariant
antisymmetric tensor. $\e^{123}$ is defined to be 1, and other
components are obtained by antisymmetry. Indices are raised and
lowered with the metric $g_{ij}$. Therefore, ${\cal E}^{ijk}=(\pm 1)
/\sqrt{g}$. $S_3$ is the part of the action with six derivatives
\be
S_3=\int dtd^3x\sqrt{g}N\left[ \beta C_{ij}C^{ij}+\beta_1 R\square
R+\beta_2R^3+\beta_3RR_{ij}R^{ij}+\beta_4
R_{ij}R^{ik}{R^{j}}_k\right].
\ee

The action with detailed balance corresponds to the following values
for the coefficients above
\be
\!\!\!\!\!\!\!\!\!\! \alpha_{DB}\!=\!
\frac{2}{\kappa^{2}}\sp \beta_{DB}\!=\!
-\frac{\kappa^{2}}{2w^{4}}\sp \gamma_{DB}\!=\!
\frac{\kappa^{2}\mu}{2w^{2}}\sp
\zeta_{DB}\!=\!-\frac{\kappa^{2}\mu^{2}}{8} \label{13}\nn\ee \be
\eta_{DB}\!=\!
\frac{\kappa^{2}\mu^{2}}{8(1\!-\!3\lambda)}\frac{1\!-\!4\lambda}{4}\sp
\xi_{DB}\!=\!
\frac{\kappa^{2}\mu^{2}}{8(1\!-\!3\lambda)}\Lambda_{W}\sp
\sigma_{DB}\!=\!
\frac{\kappa^{2}\mu^{2}}{8(1\!-\!3\lambda)}(-3\Lambda_{W}^{2})
\label{14}
\ee
as well as $\beta_1=\beta_2=\beta_3=\beta_4=0$.

The action in (\ref{12}) is invariant under a restricted class of
diffeomorphisms
\be
t'=h(t)\sp x'^{\,i}=h^i(t,\vec x). \label{16}
\ee
The transformation of the metric under infinitesimal diffeomorphisms
is
\be
\delta
g_{ij}=\pa_i\e^kg_{jk}+\pa_j\e^kg_{ik}+\e^k\pa_kg_{ij}+f\dot g_{ij}
\ee
\be
\delta N_i=\pa_i\e^jN_j+\pa_j\e^jN_i+\dot \e^j g_{ij}+\dot f
N_i+f\dot N_i\sp \delta N =\e^j\pa_j N+\dot f N+f{\dot N}.
\ee

This is the first version of the Ho\v rava-Lifshitz theory we will
analyze in this paper. A second version, called projectable, assumes
that $N$ is a function of time only and can therefore be set to one
(if non-zero) by a diffeomorphism. We will discuss spherical
solutions to this second theory in a subsequent publication.

To proceed further with our solutions we will make the simplifying assumption that $\beta_1=\beta_2=\beta_3=\beta_4=0$.
The reason is that apart from simplifying the equations, the qualitative structure of the solutions we will find is expected to be similar.
Therefore, the action we will consider is
\be
S\!=\!\!\int\!
dtd^3x\sqrt{g}N\!\left[\alpha (K_{ij}K^{ij}\!-\!\l K^2)\!+\!\beta
C_{ij}C^{ij}\!+\!\gamma {\cal E}^{ijk}R_{il}\na_j{R^l}_{k}
\!+\!\zeta R_{ij}R^{ij}\!+\!\eta R^2\!+\!\xi
R\!+\!\sigma\right]\!, \label{12}
\ee
where the various coefficients are assumed independent.

\subsection{The IR limit around flat space}

Around flat space the IR action simplifies
\be
S\to S_E=\int dtd^3x
\sqrt{g}N\left[\alpha(K_{ij}K^{ij}- \lambda K^2)+\xi
R+\sigma\right]. \label{17}
\ee
Defining $x^0=ct$, choosing
$\lambda=1$ and
\be
c=\sqrt{\xi\over \alpha}\sp M_p^2\equiv 16\pi
G=\frac{1}{\sqrt{\alpha\xi}}\sp \Lambda=-{\sigma\over 2\xi},
\label{18}
\ee
the action is that of Einstein
\be
S_E={1\over 16\pi
G}\int d^4x \sqrt{\tilde g}\left[K_{ij}K^{ij}-
K^2+R-2\Lambda\right]={1\over 16\pi G}\int d^4x \sqrt{\tilde g}
\left[\tilde R-2\Lambda\right]. \label{19}
\ee
The full space-time
metric $\tilde g_{\m\n}$ is
\be
\tilde g_{00}=-N^2+N_{i}g^{ij}N_j\sp
\tilde g_{0i}=N_i\sp \tilde g_{ij}=g_{ij}\sp \det[\tilde g]=\det[g]
N^2, \label{20}
\ee
while the inverse metric $\tilde g^{\m\n}$
\be
\tilde g^{00}=-{1\over N^2}\sp \tilde g^{0i}={N^i\over N^2}\sp
\tilde g^{ij}=g^{ij}-{N^iN^j\over N^2}. \label{21}
\ee

\section{Static, spherically symmetric solutions with zero shift}

The most general static and spherically symmetric metric
can be written as
 \be
ds^2=-(\hat{N}(r)^{2}-N_r(r)^2)f(r)dt^{2}+2N_r(r)drdt+{dr^2\over
f(r)}+r^2d\Omega_{k}^2, \label{bun1}
\ee
where $d\Omega_{k}^2$ is the
metric of a two-dimensional maximally symmetric space. For $k=1$ it
is a sphere of radius 1, for $k=-1$ it is pseudo-sphere of radius
one, and for $k=0$ it is a two-dimensional torus. We have also
parameterized the lapse $N^2=\hat N^2 f$ for further convenience.

Unlike standard general relativity, the reduced diffeomorphism
invariance here is not enough to set $N_r$ to zero. However to
simplify the equations we will consider the ansatz where $N_r=0$.
This will give as only a subset of all possible solutions. The
metrics therefore we consider are
\be
ds^2=-\hat{N}(r)^{2}f(r)dt^{2}+{dr^2\over f(r)}+r^2d\Omega_{k}^2.
\label{a2}
\ee
We now substitute this ansatz into the equations of
motion for the (source-less) action (\ref{12}). They were  first
derived in \cite{kk} and
 reproduced here in appendix  \ref{a}.

Note that since for the metrics (\ref{a2}) both $K_{ij}$ and the
Cotton tensor vanish, it is only the following part of the action
that contributes to the non-trivial equations
\be
S\!=\!\!\int\!
dtd^3x\sqrt{g}N \left[\zeta R_{ij}R^{ij}\!+\!\eta R^2\!+\!\xi
R\!+\!\sigma\right]\!, \label{a3}
\ee
Therefore, only the
coefficients $\zeta,\eta,\xi,\sigma$ enter the relevant equations.

We will assume that $\xi>0$ so that we have regular GR at weak
curvatures. Because of the existence of the couplings cubic in the
curvatures, there are no obvious constraints on the signs of
$\zeta,\eta$. Subtler constraints will be obtained later from the
behavior of non-trivial/non-linear solutions. We will be mostly interested
in  solutions that have either asymptotically AdS
or asymptotically flat behavior, although our results are valid also in other cases.
 There are cases for example where the asymptotic
behavior is dS, but in such cases the natural solutions should be
time dependent.  The asymptotic behavior is not only governed by the
``cosmological'' constant $\s$ but also by the
 curvature-squared couplings $\zeta,\eta$
via the known phenomenon of self-acceleration.

In the sequel we will use units of time so that $c=1$ which via (\ref{18}) implies $\a=\xi$.

The $N$ equation implies
\be
(3\zeta
\!+\!8\eta)r^2f'^2+4r(f-k)\left[(\zeta\!+\!4\eta)f'-\xi
r\right]-4\xi r^{3} f'+4(\zeta\!+\!2\eta)(f-k)^2+2\sigma r^4 =0.
\label{bun2}
\ee
The 12, 13 and 23 equations are trivially
satisfied. The 11 component of the spatial $(ij)$ equations gives
\be
A_{11}\hat N'+ B_{11}\hat{N}=0, \label{bun2a1} \ee where \be
A_{11}=4rf\left[(3\zeta\!+\!8\eta)rf'+2(\zeta\!+\!4\eta)(f-k)-2\xi
r^2\right] \label{tufa}
\ee
\be
B_{11}\!=\!(3\zeta+8\eta)r^{2}(4ff''+f'^{2})+4(f-k)[(\zeta+4\eta)rf'-\xi
r^{2}-2k(3\zeta+8\eta)]+2r^{3}(\sigma r-2\xi
f')-4(5\zeta+14\eta)(f-k)^{2}. \label{mufa}
\ee
\noindent The 22
equation (the 33 is identical to 22 due to symmetry) is not
independent but can be derived from (\ref{bun2}), (\ref{bun2a1}) by
differentiation.

Using (\ref{bun2}),
the expression (\ref{mufa}) assumes the simpler form
\be
B_{11}=4(3\zeta\!+\!8\eta)f\,[r^{2}f''-2(f-k)], \label{bun24}
\ee
and
equation (\ref{bun2a1}) becomes
\be
\hat{A}_{11}(\ln{\hat{N}})'+\hat{B}_{11}=0, \label{bun25}
\ee
 where
\bea &&
\hat{A}_{11}=r[(3\zeta\!+\!8\eta)rf'+2(\zeta\!+\!4\eta)(f-k)-2\xi
r^2]\label{bun26}\\
&&\hat{B}_{11}=(3\zeta\!+\!8\eta)\,[r^{2}f''-2(f-k)].\label{bun27}
\eea
Therefore, we end with two  equations (\ref{bun2}) and
(\ref{bun25}) for the two unknown functions $f(r)$, $\hat N(r)$.

In order to solve equations (\ref{bun2}), (\ref{bun25}), we will start
with appropriate  special cases in the space of parameters $\zeta, \eta$, and will then proceed
to  the general case.

\subsection{Large distance asymptotics}

In most of the solutions we will find, we have recognizable large
distance asymptotics of the form
\be
f(r)=k-{\Lambda_{eff}\over
3}r^2-{2G M\over r}+{\cal O}(r^{-4}) \sp N^2=\hat N^2
f=k-{\Lambda_{eff}\over 3}r^2-{2G\tilde M\over r}+{\cal O}(r^{-4}).
\label{t26}
\ee
 In all cases with such asymptotics $M$ and $\tilde M$ are the same.
 There are however special cases where the large distance
asymptotics above are not valid. This included the case of detailed
balance.

\subsection{On horizons and singularities}

In standard general relativity and its generalizations that keep diffeomorphism invariance intact,
horizons and singularities play an important role and their presence is relatively easy to discern.
Although there are various types of singularities from serious to mild ones (where all curvature invariants are regular)
the tools for their detection involve geodesics and curvature.

In theories without full diffeomorphism invariance, both of the previous concepts are harder to discern.
The fact that particles can have non-standard dispersion relations, and therefore no uniform maximal speed, implies that
the notion of a horizon may be different, and indeed, a concept that is probe dependent.
Indeed, in appendix \ref{geo} we analyze geodesics of particles with non-standard dispersion relations, which suggest that in
ultra-luminal cases there is no horizon in the standard sense.

A different issue is to what extend ``horizons" as defined by the divergence of metric components are indeed regular.
In diffeomorphism invariant theories, a change of coordinates suffices to make the case.
In theories with reduced diffeomorphism invariance, this may not be always the case. Indeed, several well behaved coordinate systems
involve coordinate transformations that are not symmetries.
Moreover, even coordinate systems (like the tortoise coordinate) that can be reached by defacto symmetries may be not well behaved coordinate systems.
This implies that for particles with standard dispersion relations the horizon may be a singularity.

  Finally, there is very little known about the relation of singularities with curvatures in such theories.
  Indeed, in HL theories in particular, both 3 and 4-dimensional curvatures can be computed, but as the previous
argument suggests, there may be more sources for singularities.

  It is fair to say we know little in this direction, but the problem is of fundamental importance.
  We will start its study in this paper although our results are so far modest.

\subsection{Thermodynamics\label{thermo}}

Black-hole solutions with horizons in general relativity have a well understood thermodynamic formulation
that involves the first law and the associated Bekenstein-Hawking entropy.
The mass of the black hole can be computed from the asymptotic behavior of the solution.
The temperature can be computed via the Gibbons-Hawking
technique, by rotating to Euclidean space and asking for regularity at the horizon.
Finally, the entropy can be computed from the horizon area.

In the case of general non-relativistic black holes, the notion of thermodynamics variables are not obviously well-defined.
To start with, the notion of a horizon depends on the probe particles involved. For example, if the theory contains
two types of particles with different relevant ``speeds of light", then each particle sees a different horizon.
The upshot is that entropy, if defined in terms of the horizon area, is particle-dependent.

There are more complicated cases where particles have deformed
dispersion relations and sometime no upper limit on the speed. This
indeed happens for example in the UV regime of the HL gravity
theories. In such a case the notion of the horizon seems not that
well-defined. However, in such theories the notion of a geodesic is
modified and propagation should be computed from scratch. Such
issues have been discussed in several articles
\cite{liberati}, \cite{corley}, \cite{parentani} and we leave them for
future exploration.

Not only the entropy but the notion of temperature is also probe dependent in the absence of relativistic invariance.
In \cite{parentani} the Hawking radiation spectrum was computed assuming particles have modified dispersion relations and found to deviate
  from the thermal distribution.

Finally, there is always the question of validity of the first law of thermodynamics,
but experience so far seems to indicate that it is generically valid
provided all ADM data are included.

All of the above have been linked to wither strange or unacceptable behavior
 in non-relativistic black holes. In particular, in simple cases,
it was shown that energy can be pumped out indefinitely, casting doubts on particular realizations \cite{sibiryakov}.

All such issues need to be explored in depth, but we will not do this in the present paper.
 This is why we call the spherically symmetric solutions we found
``black holes'', as it is not clear when and for whom they are
black.

In the sequel,  we will ``define" our thermodynamics by assuming
that all relevant probes have standard Lorentz-invariant dispersion
relations, as this guarantees that the temperature calculated \`a la
Gibbons-Hawking is relevant\footnote{This is obviously false with
gravitons, and such thermodynamics is inapplicable in their case.}.
 Moreover, it will identify horizons as the largest root of the function $f(r)$.
 This will coincide with the largest root of the function $N(r)$.

 We will also assume that the first law of thermodynamics is correct.
 This was shown for spherically symmetric cases in \cite{ohta} but the general case is still open.
 It is important to note that the proper definition of the thermodynamic energy is given by the mass parameter $\tilde M$ in
 (\ref{t26}). The reason is that it scales properly under constant rescaling of the time coordinate.
 Knowing the mass and temperature
allows us to compute the entropy up to an additive constant that we cannot fix from first principles.

Note that in higher derivative theories of gravity that are fully
diffeomorphism invariant we have the Wald formula for the entropy
\cite{wald}. Unfortunately, it is not at all obvious if such a
formula or its modification is applicable in our case.

\section{No quadratic curvature terms}

\noindent $\bullet$ \,\,\,$\zeta\!=\!\eta=0$. This is the simplest case,
and reduces to Einstein gravity plus cosmological constant.
 Indeed, equation (\ref{bun2}) becomes
 \be
r
f'+(f-k)-\frac{\sigma}{2\xi}r^{2}=0\;. \label{bun2a}
\ee Its
solution is
\be
f(r)=k+\frac{\sigma}{6\xi}r^{2}-\frac{c}{r}=k-{\Lambda r^2\over
3}-{2GM\over  r}\,\,, \label{bun2b}
\ee
 with  $c$ an integration constant, the mass $M={c\over 2G}$\,, and $G,\Lambda$ have been defined in (\ref{18}).
In particular, since we use units in which
 $c=1$, we have that
\be
 M_p^2\equiv 16\pi G=\frac{1}{\xi}.
 \ee

 From
(\ref{bun27}), we obtain  $\hat{B}_{11}=0$, therefore, the solution
of equation (\ref{bun25}) is $\hat{N}(r)=\hat{N}_{o}$ a constant.
Rescaling time we can set this constant to one so that
\be
\hat{N}(r)=1. \label{bun2b1}
\ee \noindent

This is the standard (A)dS-Schwarzschild solution.

\section{Special quadratic curvature combinations}

There are two possibilities that must be analyzed separately:

\subsection{\Large\bf $\zeta\!+\!3\eta=0$,
$\zeta\!\cdot\!\eta\neq 0$.\label{s1}}

In this case, equation (\ref{bun2}) becomes
\be
r^{2}f'^{\,2}+4\Big[\frac{\xi}{\eta}r^{2}\!-\!(f\!-\!k)\Big]r
f'+4\Big[\frac{\xi}{\eta}r^{2}\!+\!(f\!-\!k)\Big](f\!-\!k)-\frac{2\sigma}{\eta}r^{4}=0.
\label{t16}
\ee

We define a new function $y$ and a new coordinate $R$
\be
f(r)=k+\frac{\sigma}{6\xi}r^{2}+\frac{\xi}{\eta}r^{2}y(r)
\,\,\,\,,\,\,\,\,R=\ln{r}\,\,. \label{t15}
\ee
Equation (\ref{t16})
becomes
\be \dot{y}^{2}+4 \dot{y}+12 y=0, \label{t17}
\ee
where a
dot is a derivative with respect to $R$.

The general solution of equation (\ref{t17}) is given by the
implicit expression
\be
 (\sqrt{1\!-\!3y}- \e)\,\,e^{\e
\sqrt{1-3y}}=\Big(\frac{r_{o}}{r}\Big)^{\!3}, \label{t19}
\ee
where
$\e=\pm 1$ and $r_{o}$ is an integration constant (positive or
negative). The modified lapse is
\be
\hat{N}(r)=\frac{e^{\e(\sqrt{1-3y}-1)}}{\sqrt{1\!-\!3y}}\,\,,
\label{t18}
\ee
where we have fixed the multiplicative integration
constant appropriately. The details of the derivation are given in
appendix \ref{c}.

We will now study the large distance, $r\to\infty$ behavior of the solutions above.
\begin{itemize}

\item For large distances $r\!>\!>\!|r_{o}|$, the $\e=1$ branch of
(\ref{t19}) can be approximated as $y\sim r^{-3}$, and therefore
\be
f(r)\simeq k+{\sigma\over 6\xi}r^{2}-{2G M\over r}+{\cal
O}\left({r^{-3}}\right) \sp {G M}={1\over 3e}{\xi\over \eta}r_o^3.
\ee
For an asymptotically AdS or flat solution we must therefore
impose $\s\geq 0$.

For the modified lapse we obtain
\be
\hat N(r)\simeq 1+{r_o^6\over
2e^2 r^6}+{\cal O}\left({r^{-9}}\right).
\ee
This implies that here
$\tilde M=M$.

\item In the case with $\e=-1$, the large distance limit $r\to\infty$ is achieved when $y\to -\infty$.
We find to leading order that (for $r_{o}>0$)
\be
y(r)\simeq
-3\left[\log\left(\frac{r}{r_{o}}\right)\right]^{2}+{\cal O}(1)\sp
\hat N(r) \simeq {1\over e|\log^2(r^3/r_o^3)|}\left({r_o\over
r}\right)^{3}+\cdots\,,
\ee
from which we obtain
\be
f(r)\simeq
k+{\s\over 6\xi}r^2+{\xi r^2\over 3\eta}
\left[\left(\log\left({r^3\over
r_o^3}\right)-\log\log\left({r^3\over
r_o^3}\right)\right)^2-1\right]+\cdots \ee Expanding the lapse at
large distance we obtain \be N^2\simeq k+{\s\over 6\xi}r^2+{\xi
r^2\over 3\eta} \left[\left(\log\left({r^3\over
r_o^3}\right)-\log\log\left({r^3\over
r_o^3}\right)\right)^2-1\right]-{2G\tilde M(r)\over  r}+{\cal
O}(r^{-4}) \ee with \be {2G\tilde M(r)}={r_o^3\over
e\log^2(r^3/r_o^3)}\!\left[{\xi\over
3\eta}\!\left(\!\log\left({r^3\over
r_o^3}\right)\!-\!\log\log\left({r^3\over
r_o^3}\right)\right)^2\!-1+{\s\over 6\xi}\right]+{\cal
O}\left({1\over \log^3(r^3/r_o^3)}\right) \!.\!\!
\ee

In this case there are logarithmic corrections to the effective cosmological constant as well as the Newtonian mass.
They may be interpreted as the mass running logarithmically, and becoming smaller at large distances.
We will not consider this solution further as its interpretation is not clear.
\end{itemize}

In the above analysis we have assumed the coupling of the Einstein
term $\xi\!\neq\!0$. If we evaluate the solution for couplings
relevant for detailed balance as in (\ref{14}) with
$\lambda=\infty$, we obtain
\be
\zeta_{DB}+3\eta_{DB}=0\sp
\xi_{DB}=\sigma_{DB}=0.
\ee
Then, equation (\ref{t16}) has the
solution $f(r)=k+cr^{2}$, while (\ref{bun25}) leaves $\hat{N}$
unconstrained, which is one of the $\lambda=\infty$ solutions of
\cite{pope}.

\subsubsection{{Associated Thermodynamics }}

We abide to the assumptions of section \ref{thermo} to study the
thermodynamics of the solution above. The relevant solution is the
one with $\e=1$ with appropriate asymptotic behavior. The  black
hole horizon $r_{+}$ is defined as the largest root of the equation
$f(r)=0$, and therefore satisfies
 \be
\Big|\sqrt{1\!+\!\frac{3\eta}{\xi}\Big(\frac{k}{r_{+}^{2}}\!+\!\frac{\sigma}{6\xi}\Big)}-1\Big|
\,e^{\sqrt{1+\frac{3\eta}{\xi}(\frac{k}{r_{+}^{2}}+\frac{\sigma}{6\xi})}}=\Big(\frac{r_{o}}{r_{+}}\Big)^{3}\,.
\label{bun2g6} \ee

 The Hawking temperature is computed by the Gibbons-Hawking relation
\be
T=\frac{1}{4\pi}\sqrt{(N^{2})'f'}\,|_{r_{+}}=\frac{1}{4\pi}\hat{N}f'|_{r_{+}}
\label{ta19}
\ee
as
\be
T(r_{+})=\frac{r_{+}}{2\pi}\Big[\frac{\xi}{\eta}-
\frac{\frac{k}{r_{+}^{2}}\!+\!\frac{\xi}{\eta}}{\sqrt{1\!+\!\frac{3\eta}{\xi}
(\frac{k}{r_{+}^{2}}\!+\!\frac{\sigma}{6\xi})}}\Big]\,e^{
\sqrt{1+\frac{3\eta}{\xi}
(\frac{k}{r_{+}^{2}}+\frac{\sigma}{6\xi})}\,-1}\,\,. \label{bun2g7}
\ee

We may rewrite the mass as a function of the position of the horizon
as
\be
M(r_{+})=\Big|\frac{\xi}{\eta}\Big|\,\,\frac{r_{+}^{3}}{3eG}\,\,
\Big|\sqrt{1\!+\!\frac{3\eta}{\xi}\Big(\frac{k}{r_{+}^{2}}\!+\!\frac{\sigma}{6\xi}\Big)}-1\Big|
\,e^{\sqrt{1+\frac{3\eta}{\xi}(\frac{k}{r_{+}^{2}}+\frac{\sigma}{6\xi})}}\,\,.
\label{bun2g8}
\ee

We may now use the first law of thermodynamics to calculate  the
entropy by integrating $dS\!=\!T^{-1}dM$, or $S\!-\!S_{o}\!=\!\int\!
T^{-1}\frac{dM}{dr_{+}}dr_{+}$\,. $S_o$ is an additive constant to
the entropy that we cannot fix from first principles.

 It is
interesting to note that the entropies we will find below, whenever
they  accept the limit $r_{+}\!\rightarrow 0$ (i.e. $\frac{\eta
k}{\xi}\!>\!0$ for $k\neq 0$), in this limit the entropies  are
finite. This is in contrast to various cases studied in the recent
literature where  the entropy diverges in this limit. For example,
this is the behavior found in \cite{bh}, that studied the entropy in
the detailed balance and deformed detailed balance cases. We will
therefore choose the additive constant below so that
$S(r_{+}\!\rightarrow 0)=0$.

For $k=0$ (where we must impose  $\eta\sigma\geq -2\xi^2$ for the
existence of the solution),
 integration gives directly
 \be
S(r_{+})=(sgn\sigma)\sqrt{1\!+\!\frac{\eta\sigma}{2\xi^{2}}}\,~\frac{\pi
r_{+}^{2}}{G}. \label{bun2g8a}
\ee

For $k=\pm 1$  we have to distinguish the following cases
\begin{itemize}

\item $1\!+\!\frac{\eta\sigma}{2\xi^{2}}>0$
\be
{G\over \pi}{S(r_{+})\over sgn\Big(\!\frac{k}{r_{+}^{2}}\!+\!\frac{\sigma}{6\xi}\!\Big)}=
\,\!r_{+}^{2}\sqrt{1\!+\!\frac{3\eta}{\xi}\Big(\frac{k}{r_{+}^{2}}\!+\!\frac{\sigma}{6\xi}\Big)}
+\frac{3\eta k}{2\xi\sqrt{1\!+\!\frac{\eta\sigma}{2\xi^{2}}}}\ln{
\frac{\sqrt{1\!+\!\frac{\eta\sigma}{2\xi^{2}}}\!+\!
\sqrt{1\!+\!\frac{3\eta}{\xi}(\frac{k}{r_{+}^{2}}\!+\!\frac{\sigma}{6\xi})}}
{\Big|\sqrt{1\!+\!\frac{\eta\sigma}{2\xi^{2}}}\!-\!
\sqrt{1\!+\!\frac{3\eta}{\xi}(\frac{k}{r_{+}^{2}}\!+\!\frac{\sigma}{6\xi})}\Big|}}
 \label{bun2g9}
\ee

\item
$1\!+\!\frac{\eta\sigma}{2\xi^{2}}<0$ (where we must have  $\frac{\eta
k}{\xi}\!>\!0$)
\be
 S(r_{+})\!=\!\frac{3\eta |k|}{2\xi
G}\frac{\pi^{2}}{\sqrt{|1\!+\!\frac{\eta\sigma}{2\xi^{2}}|}}
+\frac{\pi}{G}
sgn\Big(\!\frac{k}{r_{+}^{2}}+\frac{\sigma}{6\xi}\!\Big)\!
\Bigg{\{}\!r_{+}^{2}\sqrt{1\!+\!\frac{3\eta}{\xi}\Big(\!\frac{k}{r_{+}^{2}}\!+\!\frac{\sigma}{6\xi}\!\Big)}
\,-\label{bun2g13}
\ee
$$
-\,\frac{3\eta
k}{\xi\sqrt{|\!1\!+\!\frac{\eta\sigma}{2\xi^{2}}\!|}}\arctan{ \frac{
\sqrt{1\!+\!\frac{3\eta}{\xi}(\!\frac{k}{r_{+}^{2}}\!+\!\frac{\sigma}{6\xi}\!)}}
{\sqrt{|1\!+\!\frac{\eta\sigma}{2\xi^{2}}|}}} \Bigg{\}}
$$

\item
$1\!+\!\frac{\eta\sigma}{2\xi^{2}}=0$ (where we must have $\frac{\eta
k}{\xi} \!> \!0$)
\be
{G\over \pi}{S(r_{+})\over sgn\Big(\!\frac{k}{r_{+}^{2}}\!+\!\frac{\sigma}{6\xi}\!\Big)}=
\sqrt{\frac{3\eta
k}{\xi}}\,2r_{+} \,. \label{bun2g11}
\ee

\end{itemize}

\subsection{\bf\Large $3\zeta\!+\!8\eta\!=\!0$,
$\zeta\!\cdot\!\eta\neq 0$.\label{s2}}
\vskip .6cm

Having dealt with the case $\zeta\!+\!3\eta= 0$, we may henceforth
assume that $\zeta\!+\!3\eta\neq 0$, and we will define  the
function $g(r)$ as
\be
f(r)=k+\frac{\xi}{4(\zeta\!+\!3\eta)}r^{2}+g(r). \label{bun3}
\ee
Equation (\ref{bun2}) becomes
\be
(3\zeta\!+\!8\eta)r^{2}g'^{\,2}+4(\zeta\!+\!4\eta)r g
g'+4(\zeta\!+\!2\eta)g^{2}+\frac{1}{2}\Big(\!4\sigma\!-\!\frac{3\xi^{2}}{\zeta\!+\!3\eta}\Big)r^{4}=0.
\label{bun3a}
\ee
The coefficients of equations (\ref{bun26}) and
(\ref{bun27}) simplify to
\be
\hat
A_{11}=r\left[(3\zeta\!+\!8\eta)rg'+2(\zeta\!+\!4\eta)g\right]\sp
\hat B_{11}=(3\zeta\!+\!8\eta)\left(r^2g''\!-\!2g\right).
\label{c3}
\ee

We now consider the next special case: $3\zeta\!+\!8\eta\!=\!0$,
$\zeta\!\cdot\!\eta\neq 0$.

In this case, the non-linear term in (\ref{bun3a}) vanishes and the
solution is easily found to be
\be
g^{2}=r\Big[c\!+\!\frac{2}{3\zeta}\Big(\!\sigma\!+\!\frac{6\xi^{2}}{\zeta}\!\Big)r^{3}\Big],
\label{bun4}
\ee
where $c$ is an integration constant. From
(\ref{c3}), we obtain $\hat{B}_{11}=0$, and therefore
$\hat{N}(r)=1$. A special case of this  solution was found recently in {\cite{park}},
while more special cases for this class of solutions were found in
\cite{ks}.

In order for the large distance limit to exist, we must have
\be
\frac{1}{\zeta}\left(\!\sigma+\frac{6\xi^{2}}{\zeta}\!\right)\!>\!0.
\ee
Expanding for $r\to \infty$, using (\ref{bun3}), we obtain two
asymptotic branches characterized by the sign $\e=\pm 1$
\be
f(r)\simeq k+\left(\!\e\sqrt{{2(\zeta\sigma\!+\!6\xi^{2})\over
{3\zeta^2}}}-{2\xi\over \zeta}\!\right)r^2-{2GM\over r}+{\cal
O}\left({r^{-4}}\right),
\ee
from which we obtain the mass as
\be
{GM}=-{\e c\over 4 \sqrt{{2(\zeta\sigma+6\xi^{2})\over
{3\zeta^2}}}}\,\,.
\ee
The solution with $\e=-1$ ($c>0$) is defined
for any $r$, while the one with $\e=1$ ($c<0$) is bounded from
below. In order for the solution to be asymptotically AdS when
$\e=-1$, we must have $\zeta<0$, and $\s>0$. In the opposite case
$\e=1$, we have $\zeta<0$, or $\zeta>0$ and $\s>0$.

When $\s=0$, one of the two solutions is asymptotically flat (which
one depends on the sign of $\zeta$). We obtain
\be
f(r)\simeq
k-\frac{2GM}{r}+{\cal O}\left({r^{-4}}\right)\sp {GM}=-{\zeta c\over
8\xi}
\ee
and matches the one found in \cite{ks}.

For detailed-balance with $\lambda=1$, we obtain
\be
3\zeta_{DB}+8\eta_{DB}=0\sp
\sigma_{DB}+\frac{6\xi_{DB}^{2}}{\zeta_{DB}}=0,
\ee
 therefore the solution
(\ref{bun3}) reduces to \be f(r)=k-\Lambda_{W} r^{2}+\sqrt{cr}\,,
\ee which is one of the $\lambda\!=\!1$ solutions of \cite{pope}.

 A slight
deviation from detailed-balance was defined in \cite{pope} by
\be
\zeta=\zeta_{DB}(1-\varepsilon^{2})\sp
\eta=\eta_{DB}(1-\varepsilon^{2}),
\ee
where $|\varepsilon|\ll 1$,
while $\xi, \sigma$ keep their detailed-balance values. In this
case, for $\lambda=1$, we still have $3\zeta\!+\!8\eta=0$, and
indeed
\be
\frac{1}{\zeta}\left(\!\sigma_{DB}+\frac{6\xi_{DB}^{2}}{\zeta}\!\right)
=\frac{3\Lambda_{W}^{2}}{2}\frac{\varepsilon^{2}}{(1-\varepsilon^{2})^{2}}\!>\!0,
\ee
therefore, this slight deviation is a special case of the
solution above.

\subsubsection{Associated Thermodynamics}

The thermodynamics of the solution above has been studied in detail
in several works \cite{bh}. We will therefore move forward to the generic case.

\section{The generic case}

Having disposed with the previous two special cases we may now
consider the  generic case in which
\be
(\zeta\!+\!3\eta)(3\zeta\!+\!8\eta)\neq 0.
\ee
We define the
following combinations of the relevant parameters,
\be
A=\frac{8\zeta(\zeta\!+\!3\eta)}{(3\zeta\!+\!8\eta)^{2}}\,\,,\,\,
B=\frac{1}{3\zeta\!+\!8\eta}
\Big(\!\frac{3\xi^{2}}{2(\zeta\!+\!3\eta)}\!-\!2\sigma\!\Big)\,\,,\,\,
C=\frac{16(\zeta\!+\!3\eta)}{3\zeta\!+\!8\eta}\sp A={1\over
4}C(6-C).\label{tbun7}
\ee
We define the function $g(r)$ as before
\be
f(r)=k+\frac{\xi}{4(\zeta\!+\!3\eta)}r^{2}+g(r). \label{t2}
 \ee
The differential equation (\ref{bun3a}) now becomes
\be
r^{2}g'^{\,2}+(C-4)rgg'+{1\over 2}(8-C)g^2-Br^4=0. \label{t1}
\ee

In appendix \ref{gen} we describe in detail the derivation of the
most general solution of (\ref{t1}) as well as the one for the
function $\hat N$. To describe the solutions we must distinguish two
main cases and a few subcases.

\subsection{$A>0$}

In this case, it is $B>0$ and $0<C<6$. Note that $C=0$ corresponds
to the case examined already in section \ref{s1}.

The general solution to equation (\ref{t1}) in our case is given in
implicit form as
\be
\Big(\frac{r}{r_{o}}\Big)^{\!3}
\,\Big|\!\sqrt{1\!-\!\frac{A}{B}\,\frac{g^{2}}{r^{4}}}-\e\frac{C}{2\sqrt{B}}\frac{g}{r^{2}}\Big|
=\exp\left[\frac{2\e\sqrt{A}}{C}\,\arcsin\!\Big(~\sqrt{\frac{A}{B}}\,\frac{g}{r^{2}}\!\Big)\right],
\label{t3}
\ee
where $\e=\pm 1$ is a sign and $r_o>0$ is the
integration constant. The modified lapse function is
\be
\hat{N}(r)=\sqrt{\frac{C}{6}}\,\Big(\frac{r_{o}}{r}\Big)^{\!3}\,
\frac{e^{\frac{2\e (sgng)\sqrt{A}}{C}\arcsin\!\sqrt{\frac{2A}{3C}}}}
{\sqrt{1\!-\!\frac{A}{B}\frac{g^{2}}{r^{4}}}\,\,
\Big|\!\sqrt{1\!-\!\frac{A}{B}\frac{g^{2}}{r^{4}}}-\e
\frac{C}{2\sqrt{B}}\frac{g}{r^{2}}\Big| } =\sqrt{\frac{C}{6}}\,\,
\frac{e^{-\frac{2\e\sqrt{A}}{C}\,[\,\arcsin(\sqrt{\frac{A}{B}}\,\frac{g}{r^{2}})-(sgng)\arcsin\!\sqrt{\frac{2A}{3C}}\,]}}
{\sqrt{1\!-\!\frac{A}{B}\frac{g^{2}}{r^{4}}}} \,\,, \label{t4}
\ee
where we have chosen the multiplicative integration constant so that
the function asymptotes to one at infinity distance. (\ref{t3}) in
particular implies that $|g(r)|\leq \sqrt{B\over A}r^2$.

\begin{figure}[ht]
\begin{center}
\includegraphics[width=9.5cm]{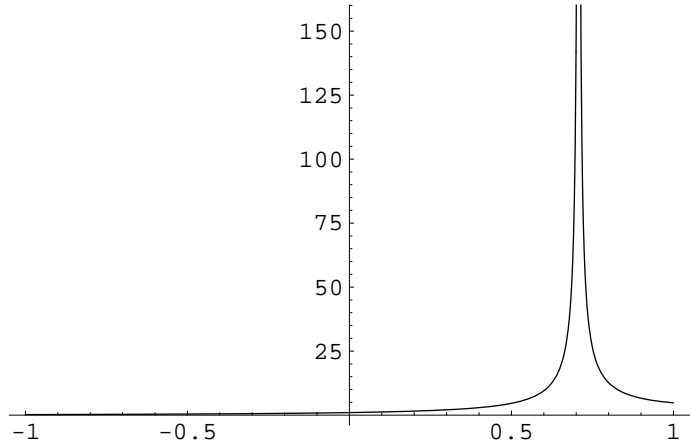}
\end{center}
\caption{${r^3\over r_o^3}$ (vertical axis) is plotted against $y=\sqrt{A\over B}{g\over r^2}$ (horizontal axis), for $\e=1$ and $C=2\sqrt{A}$.
Changing the last ratio stretches the horizontal axis, but the topology of the graph remains intact.}
\label{fig1}
\end{figure}

The plot of $(r/r_o)^3$ against  $y=\sqrt{A\over B}{g\over r^2}$ for the $\e=1$ branch
 of (\ref{t3}) is shown in figure \ref{fig1}, where we fixed $C=2\sqrt{A}$.
Note that there are two branches. On the left hand side, $r$ starts at $r_{min}=e^{-{\pi\over 2}}r_o$ (where the curvature singularity lies)
 when $y=-1$, and increases till it becomes infinite at $y={1\over \sqrt{2}}$.
 On the right hand side $y>{1\over \sqrt{2}}$, asymptotic infinity is at $y={1\over \sqrt{2}}^{+}$ and
the radius decreases until a finite value $r_{min}=e^{{\pi\over 2}}r_o$ at $y=1$.
For spherical or toroidal  symmetry, $k=0,1$, and $\zeta+3\eta>0$, the left-hand branch has a conventional
 horizon at which $f(r)$ vanishes, for a range of couplings although the right-hand one does not.
When  $\zeta+3\eta<0$ it is the right-hand branch that has a conventional horizon for a range of couplings, while the left-hand branch has $f(r)<0$.

\begin{figure}[ht]
\begin{center}
\includegraphics[width=9.5cm]{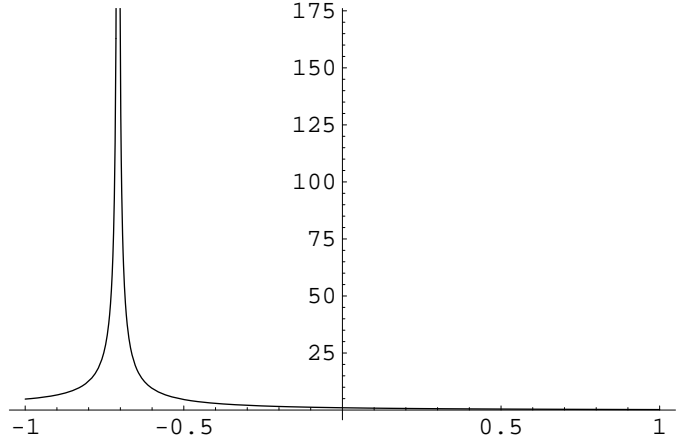}
\end{center}
\caption{${r^3\over r_o^3}$ (vertical axis) is plotted against $y=\sqrt{A\over B}{g\over r^2}$ (horizontal axis), for $\e=-1$ and $C=2\sqrt{B}$.}
\label{fig2}
\end{figure}

\begin{figure}[ht]
\begin{center}
\includegraphics[width=7cm]{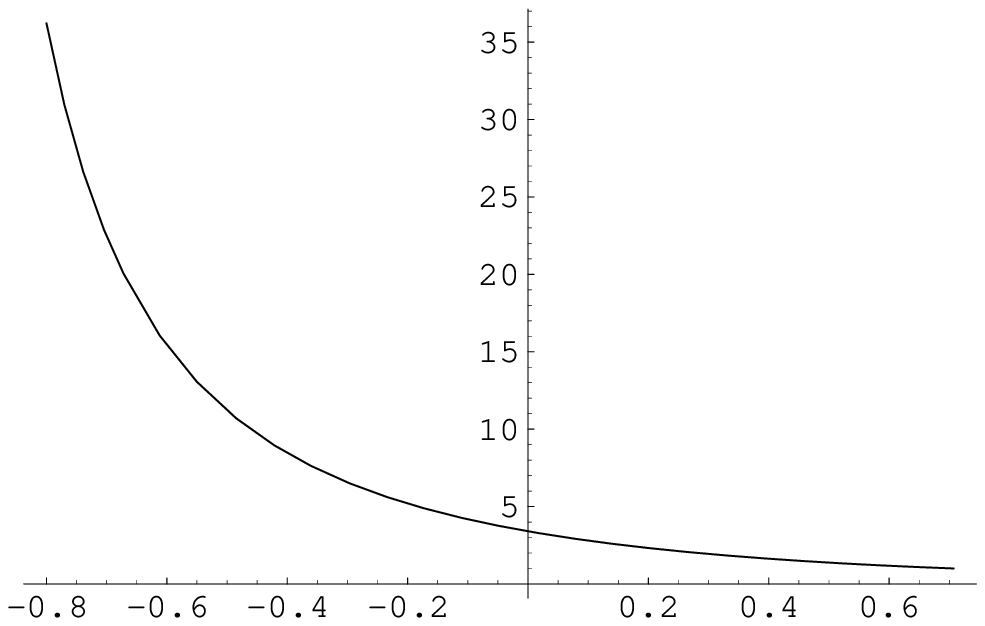}\includegraphics[width=7cm]{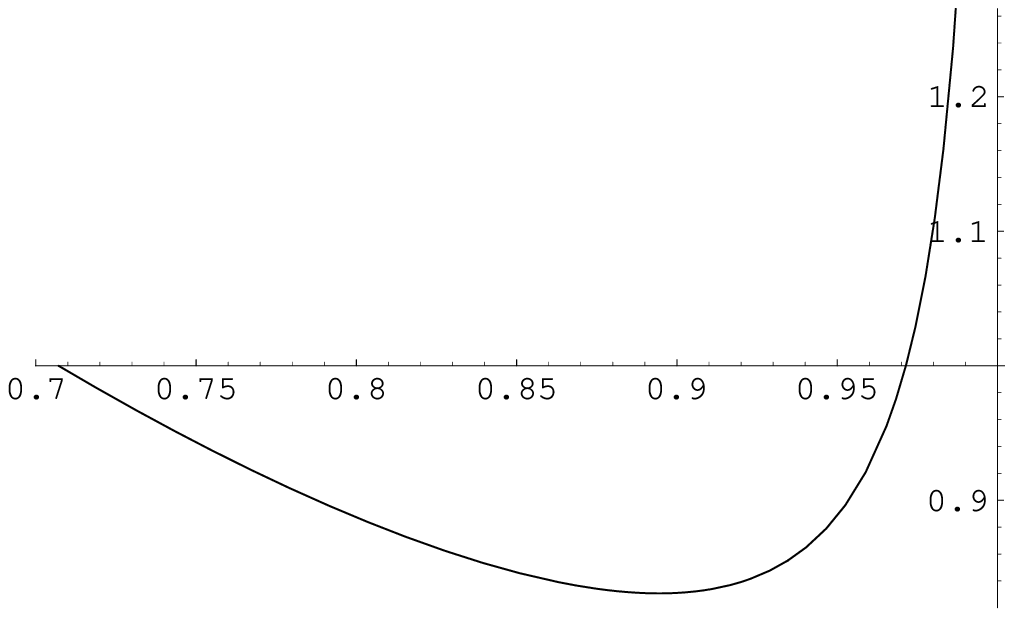}
\end{center}
\caption{On the left: the modified lapse function $\hat N$ as a function of $y$
 for the left branch of the solution in figure 1.
It asymptotes to one at $y={1\over \sqrt{2}}$ (asymptotic infinity)
and diverges from positive values at the singularity $y=-1$.
On the right: the modified lapse function $\hat N$ as a function of
$y$ for the right branch of the solution in figure 1.
It asymptotes to one at $y={1\over \sqrt{2}}$ (asymptotic infinity)
and diverges from positive values at the singularity $y=1$.
Note that unlike the previous case it is no longer monotonic.}
\label{fig3}
\end{figure}

The $\e=-1$ branch of (\ref{t3}) is shown in figure \ref{fig2} for the same values of the parameters.
It is just the mirror image of the $\e=1$ branch. For both signs of $\e$, although the solutions start at a
minimum $r_{min}$, the curvature is singular there ($y=\pm 1$).

In figure \ref{fig3} we plot $\hat{N}$ as a function of $y$ for the solutions $\e=1$ above. At the curvature singularity
 $y=\pm 1$ it is $\hat{N}\rightarrow \infty$.

For large distances $r\gg r_{o}$ (which makes sense for $\e
(sgng)>0$), the leading behavior of (\ref{t3}) for the solutions
with positive mass is
\be
 f(r)\simeq
k+\frac{1}{4(\zeta\!+\!3\eta)}\Big[\xi\!+\tilde{\e}\sqrt{\xi^{2}\!-\!\frac{4\sigma}{3}(\zeta\!+\!3\eta)}\Big]\,r^{2}
-\frac{2GM}{r}+{\cal O}(r^{-4})\,, \label{t5}
\ee with
\be
{2GM}=\frac{\sqrt{B}\,r_{o}^{3}}{3}\exp\!\Big(\frac{2\sqrt{A}}{C}\arcsin{\!\sqrt{\frac{2A}{3C}}}\Big)\,\,\,\,\,\,
,\,\,\,\,\,\, \tilde{\e}=\e\, sgn(\zeta\!+\!3\eta). \label{t6}
\ee
Moreover we have
\be
\hat{N}(r)=1+\mathcal{O}(r^{-6}). \label{t7}
\ee
This implies that $\tilde M=M$.

\subsection{$A\leq 0$}

In this case, $B$ can have either sign and $C\geq 6$ or $C<0$.

The general solution for $g(r)$ in this case is given implicitly as
\be
\Big|\!\sqrt{B\!-\!A\frac{g^{2}}{r^{4}}}-\e\frac{C}{2}\frac{g}{r^{2}}\!\Big|
\,\,\Big|\!\sqrt{B\!-\!A\frac{g^{2}}{r^{4}}}+\e\sqrt{-A}\frac{g}{r^{2}}\Big|^{\!\frac{2\sqrt{-A}}{C}}
\!=\!\Big(\frac{{r}_{o}}{r}\Big)^{3},\label{t8}
\ee
where $\e=\pm 1$
is a sign and $r_o>0$ is the integration constant. The modified
lapse function is
\be
 \hat{N}(r)=\Big(\frac{r_{o}}{r}\Big)^{\!3}
\,\,\frac{\,\sqrt{\frac{|BC|}{6}}\,\,(\sqrt{\frac{|BC|}{6}}\,|1+\frac{2\sqrt{-A}}{C}|)^{-\frac{2\sqrt{-A}}{C}}}
{\sqrt{B\!-\!A\,\frac{g^{2}}{r^{4}}}\,\,\Big|\!\sqrt{B\!-\!A\,\frac{g^{2}}{r^{4}}}-
\e\frac{C}{2}\frac{g}{r^{2}}\Big|}=
\frac{\sqrt{\frac{|BC|}{6}}\,\,\Big|\!\sqrt{B\!-\!A\,\frac{g^{2}}{r^{4}}}+\e
\sqrt{-A}\,\frac{|g|}{r^{2}}\Big|^{\frac{2\sqrt{-A}}{C}}}{(\sqrt{\frac{|BC|}{6}}\,
|1\!+\!\frac{2\sqrt{-A}}{C}|)^{\frac{2\sqrt{-A}}{C}}\,\,
\sqrt{B\!-\!A\,\frac{g^{2}}{r^{4}}}}\,\,, \label{t9}
\ee
where we
have chosen the multiplicative integration constant so that the
function asymptotes to one at infinite distance.

\begin{figure}[h]
\begin{center}
\includegraphics[width=7cm]{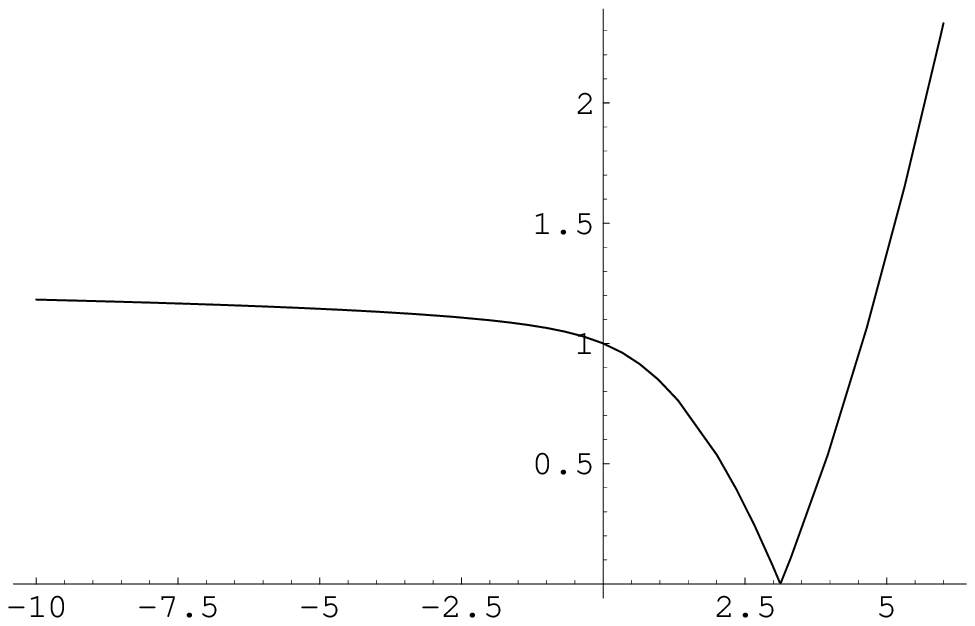} \includegraphics[width=7cm]{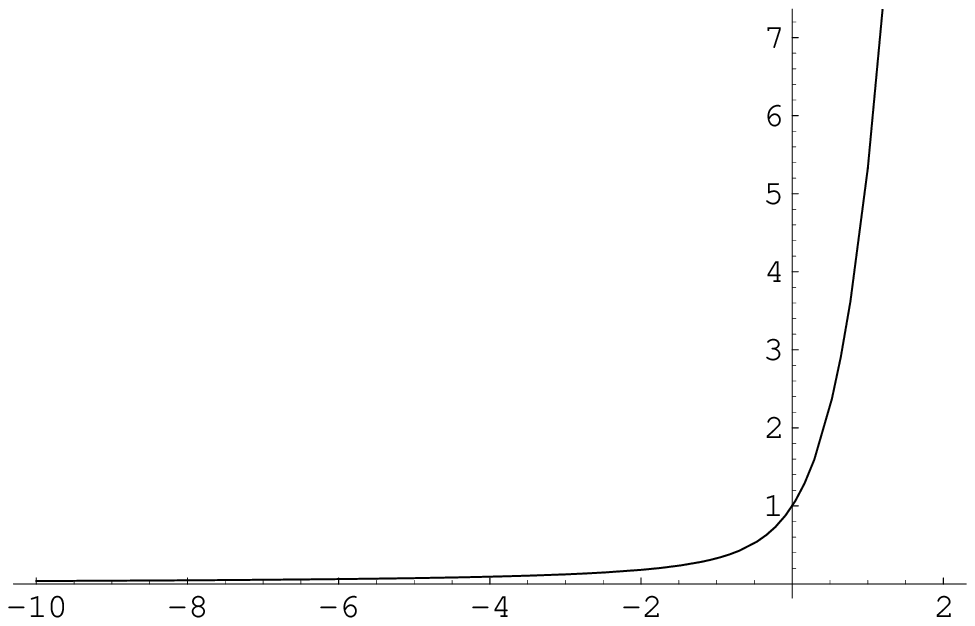}
\end{center}
\caption{On the left: $r^3_o/r^3$ (vertical axis) as a function of $y=\sqrt{|A|\over B}{g\over r^2}$
 (horizontal axis) for $\e=1$, $A<0$, $B>0$ and
${C\over 2\sqrt{B}}>1$. There are obviously two branches. The ``cusp'' corresponds to $r=\infty$.
The $r=0$ points (singularities) occur at $y=\pm \infty$.
 On the right, the same plot but with $0<{C\over 2\sqrt{B}}<1$. Here, asymptotic infinity is at $y=-\infty$, while the singularity is at $y=+\infty$.}
\label{fig4}
\end{figure}

\begin{figure}[h]
\begin{center}
\includegraphics[width=7cm]{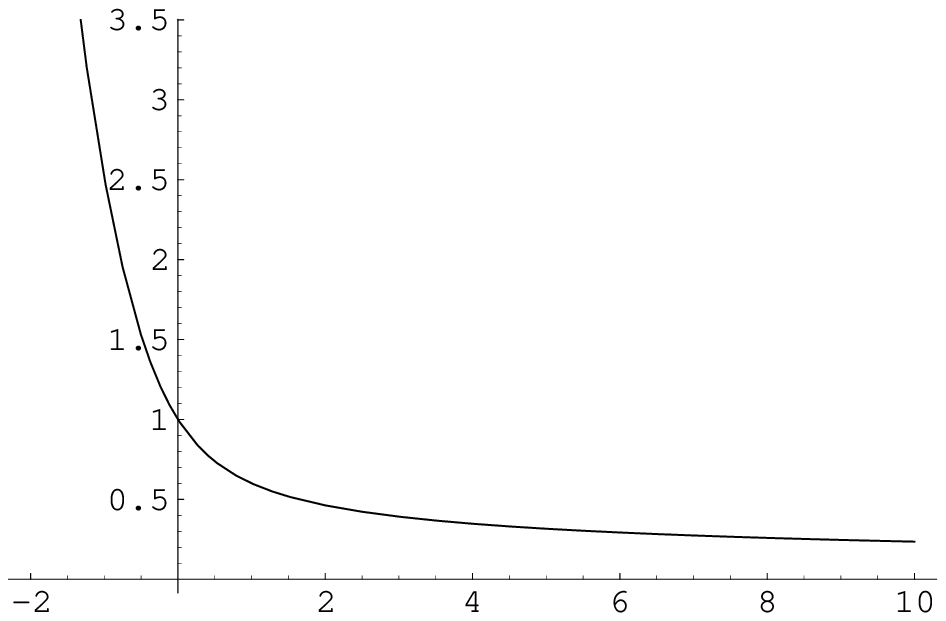} \includegraphics[width=7cm]{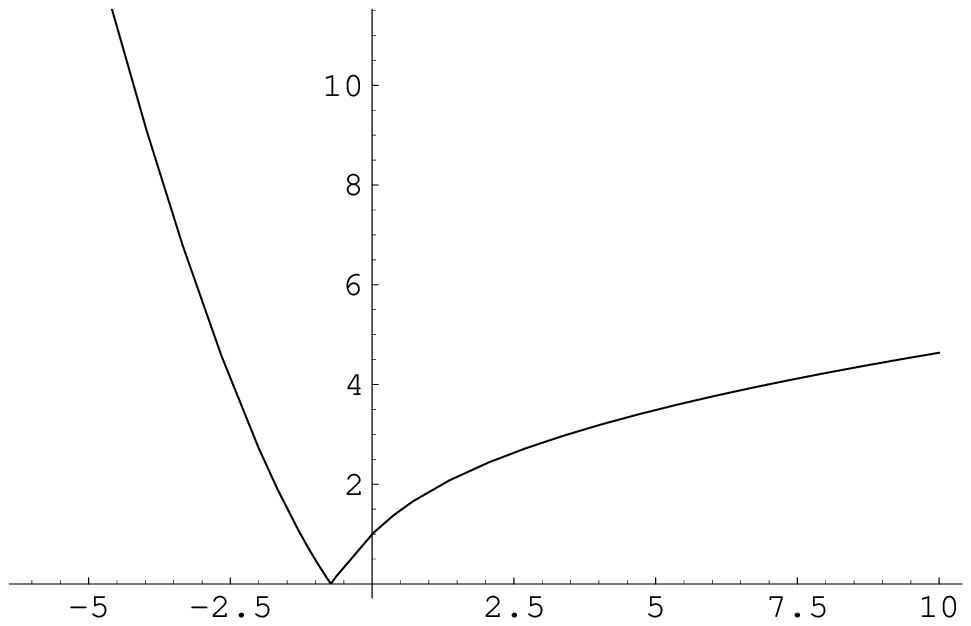}
\end{center}
\caption{On the left: $r^3_o/r^3$ (vertical axis) as a function of $y=\sqrt{|A|\over B}{g\over r^2}$
(horizontal axis) for $\e=1$, $A<0$, $B>0$ and
$0>{C\over 2\sqrt{B}}>-1$. Here, asymptotic infinity $r=\infty$ is at $y=\infty$, while the singularity $r=0$ is at $y=-\infty$.
On the right, the same plot but with ${C\over 2\sqrt{B}}<-1$.
The ``cusp'' corresponds to asymptotic infinity $r=\infty$.
The $r=0$ points (singularities) occur at $y=\pm \infty$.
 }
\label{fig5}
\end{figure}

Plots of the solutions above for different values and signs of $B$ and $C$ are portrayed in figures
\ref{fig4} and \ref{fig5}. In this case, $r$ ranges from $0$ to infinity. There is also here a curvature
singularity at $r=0$.

For large distances $r\gg r_{o}$, which for $C\geq 6$ makes sense
for $B>0$, $\e (sgng)>0$, while for $C<0$, $B<0$, $\e (sgng)<0$, the
leading behavior is achieved with the vanishing of the first factor
in (\ref{t8}) and we obtain for the solutions with positive mass

 \be
 f(r)\simeq
k+\frac{1}{4(\zeta\!+\!3\eta)}\Big[\xi\!+\tilde{\e}\sqrt{\xi^{2}\!-\!\frac{4\sigma}{3}(\zeta\!+\!3\eta)}\Big]\,r^{2}
-\frac{2GM}{r}+{\cal O}(r^{-4})\,, \label{t10} \ee \be 2GM=
\frac{r_o^3}{ 3}\Big[\sqrt{BC\over 6}\,\Big|1\!+\!{2\sqrt{|A|}\over
C}\Big|\Big]^{-\frac{2\sqrt{|A|}}{ C}}\,\,\,\,\,\, ,\,\,\,\,\,\,
\tilde{\e}=\e\, sgn(\zeta\!+\!3\eta)\,\,,
 \label{t11}
\ee
\be
\hat N(r)\simeq 1+{\cal O}(r^{-6})\,. \label{t27}
\ee
Note that
to obtain the result (\ref{t27}) one has to consider in the
expansion of (\ref{t10}) also the exact $r^{-4}$ term.
\newline
This implies that also here $\tilde M=M$.

For $C<0$, there is another way to make something vanish on the
left-hand side of (\ref{t8}) and this is to have $g/r^2\to \infty$.
This gives a different asymptotic behavior at large $r$
\be
f(r)\simeq k+\frac{1}{4(\zeta\!+\!3\eta)}\xi
r^2+{(2\sqrt{\aa})^{2\sqrt{\aa}\over \cc-2\sqrt{\aa}}~r^2\over
\left(\sqrt{\aa}+{\cc\over 2}\right)^{\cc\over \cc-2\sqrt{\aa}}}~
\left({r_o\over r}\right)^{3\cc\over \cc-2\sqrt{\aa}}+\cdots\sp \hat
N(r)\sim \left({r_o\over r}\right)^{-3{|C|+2\sqrt{|A|}\over
|C|-2\sqrt{|A|}}}\,.
 \label{t20}
\ee
Note than in this case the exponent $3\cc\over \cc-2\sqrt{\aa}$ is
negative and spans all real negative values as $-\infty<C<0$. This
is a case where the long distance behavior in the $f(r)$ function is
stronger than the generic ${\cal O}(r^2)$ AdS behavior.

\subsection{$B=0$}

The value $B=0$ is a special one for equation (\ref{t1}) with
corresponding solution
\be
g(r)=c_{1}\,r^{2-{C\over
2}+\varepsilon\sqrt{|A|}}\sp \hat{N}(r)=\hat N_o\,
r^{\frac{C}{2}-2\varepsilon \sqrt{|A|}\,+2\frac{|A|}{C}},
\label{t12}
\ee
where $\varepsilon=\pm 1$ and $c_{1},\hat{N}_{o}$ are
integration constants. Note that $B=0$ is allowed for the solution
(\ref{t8}) but not for the solution (\ref{t3}).

Detailed-balance with arbitrary $\lambda$ has
\be
A_{DB}=2{1\!-\!3\lambda\over (1\!-\!\lambda)^{2}}\sp  B_{DB}=0\sp
C_{DB}={4\over 1\!-\!\lambda} \label{t13}
\ee
and therefore
corresponds to this last case. Substituting these values for $A,C$
in (\ref{t12}) we obtain
\be
g(r)\sim
r^{(2\lambda+\varepsilon\sqrt{6\lambda-2}\,)/(\lambda-1)}\sp
\hat{N}(r)\sim
r^{-(1+3\lambda+2\varepsilon\sqrt{6\lambda-2}\,)/(\lambda-1)},\label{t14}
\ee
which is the first solution found in \cite{pope} with
generic $\lambda>1/3$.

When we choose further for the integration constant $c_1=0$ in
(\ref{t12})
 we find $g(r)=0$, and then, for detailed-balance with
arbitrary $\lambda$, the lapse $\hat{N}$ remains undetermined, which
is the second solution found in \cite{pope} with generic $\lambda$.

We end by noting that in the generic case with the exception of
$B=0$, there are always solutions with the standard asymptotic
behavior exemplified in (\ref{t5}), (\ref{t7}) (or the same
(\ref{t10}), (\ref{t27})). The solutions are generically
asymptotically AdS/dS, with a standard $1/r$ tale that defines the
mass of the solution. In all such cases the solutions come in a pair
with different asymptotic effective cosmological constants
\be
\Lambda_{eff}=-\frac{3}{4(\zeta\!+\!3\eta)}\Big[\xi\!+\tilde{\e}\sqrt{\xi^{2}\!-\!\frac{4\sigma}{3}(\zeta\!+\!3\eta)}\Big]\sp
\tilde{\e}=\pm 1.
 \label{t21}
\ee
Not only the bare cosmological constant is responsible for $\Lambda_{eff}$ but also the curvature square couplings $\zeta,\eta$.
This is special example of what is known in cosmology as self-acceleration: the generation of acceleration (background curvature) by gravity alone.

The effective cosmological constant can become zero and the solution
asymptotically flat when $\s=0$. In this case, the $\tilde{\e}=-1$
solution is asymptotically flat, while the $\tilde{\e}=1$ solution
has
\be
\Lambda_{eff}=-\frac{3\xi}{2(\zeta\!+\!3\eta)}.
 \label{t22}
\ee

\section{The horizons in the generic case}

We define again the exterior horizon position $r_+$ as the largest
root of equation $f(r)=0$. This implies that
\be
{g(r_+)\over
r_+^2}=-{k\over r_+^2}-\frac{\xi}{4(\zeta\!+\!3\eta)}=-X,
 \label{t23}
\ee
where we defined the variable $X$ that will be useful later on.

\subsection{$A>0$}
From (\ref{t3}) we obtain an equation for $r_+$
\be
\Big|\sqrt{1\!-\!\frac{A}{B}X^{\!2}} +\e\frac{C}{2\sqrt{B}}X\Big| =
\Big(\frac{r_{o}}{r_{+}}\Big)^{\!3}\,\,
e^{\frac{-2\e\sqrt{A}}{C}\arcsin(\sqrt{\!\frac{A}{B}}\,X)}.
\label{xbun20}
\ee
The existence of the square root implies an upper
bound on $r_+$
\be
\Big|\frac{k}{r_{+}^{2}}\!+\!\frac{\xi}{4(\zeta\!+\!3\eta)}\Big|\leq
\sqrt{B\over A}\,\,.
 \label{t24}
\ee
For a flat horizon $k=0$, the inequality constraints the couplings
\be
\Big(\!1+3\frac{\eta}{\zeta}\Big)\,[2\xi^{2}-\sigma(3\zeta\!+\!8\eta)]\geq
0.
 \label{t25}\ee

\subsection{$A\leq 0$}
From (\ref{t8}) we obtain an equation for $r_+$
\be
\Big|\sqrt{\!B\!-\!AX^{2}}+\e
\frac{C}{2}X\Big|\,\,|\sqrt{\!B\!-\!AX^{2}}-\e
\sqrt{\!-A}\,X|^{\frac{2\sqrt{-A}}{C}}
=\Big(\frac{r_{o}}{r_{+}}\Big)^{\!3} . \label{ybun23a}
\ee

\section{Thermodynamics in the generic case}

 We now investigate the relevant thermodynamics in the generic case following the remarks and assumptions of section
 \ref{thermo}.

\subsection{$A>0$}
As showed before, there are solutions with the standard $1/r$
asymptotic behavior (\ref{t5}),(\ref{t7}). For simplicity we assume
that $\zeta\!+\!3\eta>0$, so that $\tilde{\e}=\e$.

 The Hawking temperature is computed from (\ref{t19}) as
  \be
T(r_{+})=\frac{r_{+}}{4\pi}\sqrt{\frac{C}{6}}\,\Big[\frac{\frac{\xi
C}{8(\zeta+3\eta)}+\frac{C-4}{2}\frac{k}{r_{+}^{2}}}{\sqrt{1\!-\!\frac{A}{B}X^{2}}}
+\e\sqrt{B}\,\Big]\,e^{\frac{2\sqrt{A}}{C}\,[\,\arcsin{\!\sqrt{\!\frac{2A}{3C}}}
\,\,+\,\e\,\arcsin{(\sqrt{\!\frac{A}{B}}\,X)} \,]}\,\,,
\label{bun21}
\ee
 while the  mass from (\ref{t6}) is
\be
M(r_{+})\!=\!\frac{\sqrt{\!A}}{6G}r_{+}^{3}\Bigg[\frac{C}{2\sqrt{\!A}}X+\e
\sqrt{\frac{B}{A}\!-\!X^{2}}\,\Bigg]\,e^{\frac{2\sqrt{A}}{C}\,[\,\arcsin{\!\sqrt{\!\frac{2A}{3C}}}
\,\,+\,\e\,\arcsin{(\sqrt{\!\frac{A}{B}}\,X)} \,]}\,.
\label{bun22}
\ee

We calculate the entropy by assuming  the first law $dS=T^{-1}dM$, from which
\be
S=S_{o}+\int T^{-1}\frac{dM}{dr_{+}}dr_{+}.
\ee

For $k=0$,
integration gives
\be
S=\sqrt{\frac{6}{C}}\,\sqrt{1\!-\!\frac{A}{B}\frac{\xi^{2}}{16(\zeta\!+\!3\eta)^{2}}}\,\frac{\pi
r_{+}^{2}}{G}, \label{bun22a}
\ee
where we have set $S_{o}=0$ so that $S(r_{+}\rightarrow 0)=0$.

For $k\neq 0$ (where the inequality (\ref{t24}) is relevant)
\bea
 && \!\!\!\!\!\!\!\!\!\!\!\!\!\!\!\!\!\!\!\!\!\!\!\!
S(r_{+})\!-\!S_{o}=\frac{\pi
k}{G}\sqrt{\!\frac{6A}{BC}}\,\Bigg{[}kr_{+}^{2}\sqrt{\frac{B}{A}\!-\!
X^{2}}+\arcsin\!\Big(\sqrt{\!\frac{A}{B}}X\!\Big)\nn\\
&&\,\,\,\,\,\,\,\,\,\,\,\,\,\,\,\,\,\,\,\,\,\,\,\,\,\,\,\,\,\,\,\,\,\,\,\,\,\,\,\,\,\,\,+\,\frac{2
sgn(1\!-\!\textrm{a}^{2})}{\sqrt{|1\!-\!\textrm{a}^{2}|}}\arctan(\textrm{h})
\frac{\frac{4(\zeta+3\eta)}{\xi}\sqrt{\!\frac{B}{A}}\,\frac{k}{r_{+}^{2}}
\!+\!\sqrt{\frac{B}{A}\!-\!X^{2}}}{\sqrt{|1\!-\!\textrm{a}^{2}|}\,
X}\,\Bigg{]}, \label{bun23}
\eea
where
\be
\textrm{a}=\e
\frac{4(\zeta+3\eta)}{\xi}\sqrt{\!\frac{B}{A}}\,,
\ee
while
$\arctan$, $\arctan$$\textrm{h}$ are used for $|\textrm{a}|<1$,
$|\textrm{a}|>1$ respectively.

\subsection{$A\leq 0$}

Also here, there are solutions with the standard $1/r$ asymptotic
behavior (\ref{t10}),(\ref{t27}). We will study the case with $C\geq
6$ and for simplicity we assume that $\zeta\!+\!3\eta>0$, so that
$\tilde{\e}=\e$.

 The Hawking temperature in this case is computed to be
  \be
T(r_{+})=\frac{\hat{N}_{o}\,r_{+}}{4\pi}\,\Big[\frac{\frac{\xi
C}{8(\zeta+3\eta)}+\frac{C-4}{2}\frac{k}{r_{+}^{2}}}{\sqrt{1\!-\!\frac{A}{B}X^{2}}}
+\e\sqrt{B}\,\Big]\Bigg[\frac{\sqrt{\! X^{2}\!-\!\frac{B}{A}}-\e X}
{\sqrt{\!X^{2}\!-\!\frac{B}{A}}+\e
X}\Bigg]^{\!\frac{\sqrt{-A}}{C}}\,, \label{bun23b}
\ee
while the
mass is
\be
M(r_{+})\!=\frac{\hat{N}_{o}}{G}\,\sqrt{\!\frac{|A|}{6C}}\,r_{+}^{3}
\!\Bigg[\!\frac{C}{2\sqrt{\!|A|}}X
\!+\e\sqrt{\!X^{2}\!-\!\frac{B}{A}}\Bigg] \Bigg[\frac{\sqrt{\!
X^{2}\!-\!\frac{B}{A}}-\e X} {\sqrt{\!X^{2}\!-\!\frac{B}{A}}+\e
X}\Bigg]^{\!\frac{\sqrt{-A}}{C}}\!\!\!.\nn\\
\label{bun23c}
\ee

Using the first law, we calculate the entropy by integration as before.
For $k=0$, this integration gives again (\ref{bun22a}).

For $k\neq 0$, we obtain
 \bea
 S\!-\!S_{o}\!=\!\frac{\pi
k}{G}\sqrt{\frac{6|A|}{BC}}\,\,\Bigg\{\!\!\!&&\!kr_{+}^{2}\sqrt{\!X^{2}\!-\!\frac{B}{A}}\nn\\
&&-\,\e\log\!\Bigg[\! \frac{\sqrt{\!X^{2}\!-\!\frac{B}{A}}+\e
X\!+\!\sqrt{\!\frac{B}{|A|}}}{\sqrt{\!X^{2}\!-\!\frac{B}{A}}-\e
X\!+\!\sqrt{\!\frac{B}{|A|}}}\,\,\,\Bigg|\!\frac{\sqrt{\!X^{2}\!-\!\frac{B}{A}}+\e
(\textrm{a}\!-\!\sqrt{1\!+\!\textrm{a}^{2}})X\!-\!\sqrt{\!\frac{B}{|A|}}}
{\sqrt{\!X^{2}\!-\!\frac{B}{A}}+\e
(\textrm{a}\!+\!\sqrt{1\!+\!\textrm{a}^{2}})X\!-\!\sqrt{\!\frac{B}{|A|}}}\Bigg|
^{\frac{1}{\sqrt{1+\textrm{a}^{2}}}}\Bigg]\Bigg\},
\nn\\
\!\!\!\!\label{bun244}
\eea
where
\be
\textrm{a}=\e
\frac{4(\zeta+3\eta)}{\xi}\sqrt{\frac{B}{|A|}}\,\,.
\ee
 We note that
the expression (\ref{bun244}) has a logarithmic divergence when
$r_{+}\rightarrow 0$ for both branches $\e$, due to the first factor
inside the logarithm.

\section{Discussion and further directions}

We have found  the most general spherically symmetric solution to Ho\v rava-Lifshitz type
of gravity theories with zero shift.
The action used was general except at the level cubic in curvatures where the Cotton$^2$ combination was used.
 We do not expect new qualitatively different effects if we allow also the most general (curvature)$^3$
 couplings.

$\bullet$ It is a generic feature of the solutions found that they have
regular large distance asymptotics that are asymptotically AdS/dS or
flat. Moreover, generically the next correction is compatible with a
standard Newton's law.

There are exceptions to this results that are special. In
particular, the detailed balance action first written down by Hor\v
ava is one of these notable exceptions. As shown already first in
\cite{pope} it does not reproduce the correct Newton law at large
distances. This is part of a class of special cases analyzed here,
occurring when
\be
B=\frac{1}{3\zeta\!+\!8\eta}
\Big(\!\frac{3\xi^{2}}{2(\zeta\!+\!3\eta)}\!-\!2\sigma\!\Big)=0 ~~~{\rm or}~~~
C=\frac{16(\zeta\!+\!3\eta)}{3\zeta\!+\!8\eta}<0.
\ee
although in the second case it is not the generic solution.

$\bullet$ In one of the two general categories of parameters ($A\leq 0$)
where the horizon distance is permitted to shrink to zero, the
entropy has a logarithmic divergence in this limit. In the other general category
($A>0$) the horizon distance is bounded from below.

$\bullet$ For special values of the curvature-squared couplings
$\zeta+3\eta=0$, when the horizon position  is allowed  to go to zero
the entropy is regular in this limit.

$\bullet$ For the same values of the curvature-squared couplings $\zeta\!+\!3\eta=0$,
$\zeta\!\cdot\!\eta\neq 0$, there exists a solution with logarithmically corrected large distance asymptotics.
Its ``mass parameter" defined in the naive way, depends logarithmically on radial distance and becoming smaller at larger distances.

$\bullet$ When the cosmological constant term in the action is absent, the
effective cosmological constant of $(A)dS$ curvature is also zero
(for some branches).

$\bullet$ We have studied  geodesics of particles with finite and infinite light speeds in spherically symmetric backgrounds
with traditional horizons. We parameterize the dispersion relations of particles as $p_0^2=(\vec p^{\,2})^n$, with $n\geq 1$.
We find that when $n>1$ the traditional behavior of the horizon disappears suggesting that for such particles that black hole is
effectively ``naked".

There are several interesting questions remaining open.

\begin{itemize}

\item The meaning of the notion of a black-hole and that of horizon in a theory with particles that have modified dispersion relations
must be investigated. In particular, apparent puzzles/instabilities that seem
 to exist in such cases must be resolved or such theories will be physically discredited.
A key element that may turn out to be important is a strong downgrading of energy in cases of modified dispersion relations that
together with low energy Lorentz invariance may be a way to resolve puzzles.

\item Finer properties of the black hole found should be studied as well as extremal limit in search for stronger constraints
on the physical acceptability of such solutions

\item Spherically symmetric solutions  in the projectable theory need to be studied as no non-trivial one is known so far.

\item The study of the non-linear structure of spherically symmetric solutions in the modified non-projectable theory of \cite{blas2}
must be studied in order to evaluate the stability claims of such solutions at the non-linear level.

\end{itemize}

We plan to return to these question in the near future.

\vspace{.7 in}
\addcontentsline{toc}{section}{Acknowledgments}

\noindent {\bf Acknowledgements} \newline

We are grateful to C. Charmousis, R. Parentani  and H. B. Zhang for
valuable discussions.

 This work was  partially supported by  a European Union grant FP7-REGPOT-2008-1-CreteHEP
 Cosmo-228644, and a CNRS PICS grant \# 4172.

\newpage
\appendix
\renewcommand{\theequation}{\thesection.\arabic{equation}}
\addcontentsline{toc}{section}{APPENDIX\label{app}}
\section*{APPENDIX}
\section{The classical equations of motion\label{a}}

We now add the action of matter
\be
 S_{M}=\int d^3xdt
\sqrt{g}N~{\cal L}_{\rm matter}(N,N_i,g_{ij}) \label{30}
\ee
to the gravitational action (\ref{12}) and we will vary with respect to the
gravitational fields to obtain the equations of motion.

The equation obtained by varying N  is
\be
-\alpha\left(K_{ij}K^{ij}-\l K^2\right)+\beta C_{ij}C^{ij}+\gamma
{\cal E}^{ijk}R_{il}\na_j{R^l}_{k} +\zeta R_{ij}R^{ij}+\eta R^2+\xi
R+\sigma=J_N, \label{31}
\ee
with
\be J_N=-{\cal L}_{\rm
matter}-N{\delta {\cal L}_{\rm matter}\over \delta N}.
\ee

The equation  obtained by varying  $N_i$  is
\be
2\alpha(\na_jK^{ji}-\l\na^i K)+N{\delta {\cal L}_{\rm matter}\over
\delta N_i}=0. \label{32}
\ee

Finally, the equation of motion obtained by varying $g_{ij}$ is more
voluminous

\bea
&&\!\!\!\!\!\!\!\!\frac{1}{2}
\Big[({\cal E}^{mk\ell}Q_{mi})_{;kj\ell}\!+\!({\cal E}^{mk\ell}Q^{n}_{m})_{;kin}g_{j\ell}\!-\!
({\cal E}^{mk\ell}Q_{mi})_{;kn}^{\,\,\,\,\,\,\,;n}g_{j\ell}\!-\!({\cal E}^{mk\ell}Q_{mi})_{;k}R_{j\ell}\nn\\
&&\!\!\!\!\!\!\!\!\!\!\!
-({\cal E}^{mk\ell}Q_{mi}R^{n}_{k})_{;n}g_{j\ell}\!+\!({\cal E}^{mk\ell}Q^{n}_{m}R_{ki})_{;n}g_{j\ell}\!+\!
\frac{1}{2}({\cal E}^{mk\ell}R^{n}_{\,\,\,pk\ell}Q^{p}_{m})_{;n}g_{ij}\!-\!Q_{k\ell}C^{k\ell}g_{ij}\!+\nn\\
&&\!\!\!\!\!\!\!\!\!\!\!
{\cal E}^{mk\ell}Q_{mi}R_{j\ell;k}\Big]+\Box[N(2\eta
R\!+\!\xi)]g_{ij}\!+\!N(2\eta R\!+\!\xi)R_{ij}\!+\!2N(\zeta
R_{ik}R^{k}_{j}\!-\!\beta C_{ik}C^{k}_{j})\!\nn\\
&&\!\!\!\!\!\!\!\!\!\!\!-[N(2\eta R\!+\!\xi)]_{;ij}\!+\!\Box[N(\zeta
R_{ij}\!+\!\frac{\gamma}{2}C_{ij})]-2[N(\zeta
R_{ik}\!+\!\frac{\gamma}{2}C_{ik})]_{;j}^{\,\,\,\,;k}\!+\![N(\zeta
R^{k\ell}\!+\!\frac{\gamma}{2}C^{k\ell})]_{;k\ell}g_{ij}\!\nn\\
&&\!\!\!\!\!\!\!\!\!-\frac{N}{2}(\beta C_{k\ell}C^{k\ell}\!+\!\gamma
R_{k\ell}C^{k\ell}\!+\!\zeta R_{k\ell}R^{k\ell}\!+\!\eta
R^{2}\!+\!\xi
R\!+\!\sigma)g_{ij}\!+2\alpha\!N(K_{ik}K^{k}_{j}\!-\!\lambda
KK_{ij})\!\nn\\
&& -\frac{\alpha N}{2}(K_{k\ell}K^{k\ell}\!-\!\lambda
K^{2})g_{ij}\!+\!\frac{\alpha}{\sqrt{\!g}}g_{ik}g_{j\ell}{\partial\over
\partial t}[\sqrt{\!g}(K^{k\ell}\!-\!\lambda
Kg^{k\ell})]+\alpha[(K_{ik}\!-\!\lambda K g_{ik})N_{j}]^{;k}\!\nn\\
&& +\alpha[(K_{jk}\!-\!\lambda K
g_{jk})N_{i}]^{;k}\!-\!\alpha[(K_{ij}\!-\!\lambda K
g_{ij})N_{k}]^{;k}+(i\leftrightarrow j)=-2N{\delta {\cal L}_{\rm
matter}\over \delta g^{ij}}, \label{33}
\eea
\noindent where ${\cal
E}^{ijk}$ was defined below equation  (\ref{i12}), $;$ stands for
covariant differentiation with respect to the metric $g_{ij}$,  and
\be
Q_{ij}\equiv N(\gamma R_{ij}\!+\!2\beta C_{ij}). \label{34}
\ee

\section{ Derivation of  solution when \bf $\zeta\!+\!3\eta=0$,
$\zeta\!\cdot\!\eta\neq 0$.\label{c}}

In this case, equation (\ref{bun2}) becomes
\be
r^{2}f'^{\,2}+4\Big[\frac{\xi}{\eta}r^{2}\!-\!(f\!-\!k)\Big]r
f'+4\Big[\frac{\xi}{\eta}r^{2}\!+\!(f\!-\!k)\Big](f\!-\!k)-\frac{2\sigma}{\eta}r^{4}=0.
\label{abun2d}
\ee
We define a new function $y$ and a new coordinate
$R$
\be
f(r)=k+\frac{\sigma}{6\xi}r^{2}+\frac{\xi}{\eta}r^{2}y(r)
\,\,\,\,,\,\,\,\,R=\ln{r}\,\,. \label{abun2e}
\ee
Equation
(\ref{abun2d}) becomes
\be
\dot{y}^{2}+4 \dot{y}+12 y=0,
\label{abun2f}
\ee
where a dot is a derivative with respect to $R$.

The general solution of equation (\ref{abun2f}) must have
$y(r)<1/3$, and has two branches with $\e=\pm 1$. It is given by the
implicit expression
\be
 (\sqrt{1\!-\!3y}- \e)\,\,e^{\e
\sqrt{1-3y}}=\Big(\frac{r_{o}}{r}\Big)^{\!3}, \label{abun2g}
\ee
where $r_{o}$ is integration constant (positive or negative).

We proceed with the integration of equation (\ref{bun25}) and define
\be
Y\equiv\frac{y}{\dot{y}}={y\over
2\e(\sqrt{1-3y}-\e)}=-{\sqrt{1-3y}+\e\over 6\e}={y\over
2\e}\Big(\frac{r_{o}}{r}\Big)^{\!\!-3}e^{\e \sqrt{1-3y}},
\ee
where
in the last equality we used equation (\ref{abun2g}).

Moreover, using (\ref{abun2f}), we obtain $y=-4Y(1+3Y)$, and
therefore
\be
\dot{Y}={1+3Y\over 1+6Y}.
\ee
Equation (\ref{bun25})
now becomes
\be
\frac{d\ln\hat{N}}{dY}=-12\frac{1+3Y}{1+6Y}\,\,.
\label{abun2g1}
\ee
Equation (\ref{abun2g1}) can  be integrated to
\be
\hat{N}\!={\hat{N}_{o}\over e^{6Y\!}|1\!+\!6Y|},
\ee
 with
$\hat{N}_{o}$ integration constant. In terms of the  $y$ variable it
becomes,
\be
\hat{N}(r)=\frac{e^{\e(\sqrt{1-3y}-1)}}{\sqrt{1\!-\!3y}}\,\,,
\label{abun2g2}
\ee
 where we fixed the integration constant $\hat N_o=e^{-\e-1}$.

 Equation (\ref{abun2g2}), together with
(\ref{abun2g}) provide the general solution when
$\zeta\!+\!3\eta\!=\!0$. This solution can be given alternatively in
terms of the variable $Y$ as
 \be
f(r)=k+\frac{\sigma}{6\xi}r^{2}-\frac{4\xi}{\eta}r^{2}\,Y(1\!+\!3Y)\label{abun2g3}
\ee
\be
(1\!+\!3Y)\,\,e^{
-6Y}=\frac{e}{2}\Big(\frac{r_{o}}{r}\Big)^{\!3}\sp
\hat{N}(r)=\frac{\hat{N}_{o}}{e^{6Y}|1\!+\!6Y|}\label{abun2g5}\,\,.
\ee
Above, the branch with $\e=1$ corresponds to $Y<-1/6$, while
$\e=-1$ corresponds to $Y>-1/6$.

We will now study the large distance, $r\to\infty$ behavior of the solutions above.
\begin{itemize}

\item For large distances $r\!>\!>\!|r_{o}|$, the $\e=1$ branch of
(\ref{abun2g}) can be approximated as $y\sim r^{-3}$, and therefore
\be f(r)\simeq k+{\sigma\over 6\xi}r^{2}-{2GM\over r}+{\cal
O}({r^{-3}}) \sp GM=\frac{1}{3\e}\frac{\xi}{\eta}r_o^3. \ee For an
asymptotically AdS or flat solution we must therefore impose $\s\geq
0$.

For the modified lapse and the lapse we obtain \be \hat N(r)\simeq
1+{\cal O}({ r^{-6}})\sp N^{2}(r)\simeq k+{\sigma\over
6\xi}r^{2}-{2GM\over r}+{\cal O}({r^{-3}}).\ee

\item In the case with $\e=-1$, the large distance limit $r\to\infty$ is achieved when $y\to -\infty$.
We find to leading order that \be \sqrt{1-3y}\simeq
\log\left({r^3\over r_o^3}\right)-\log\log\left({r^3\over
r_o^3}\right)+\cdots\sp \hat N(r) \sim {1\over
|\log^{2}(r/r_o)|}\left({r_o\over r}\right)^{3}+\cdots \ee
Therefore, such terms logarithmically dominate the cosmological
constant  in the large distance limit.
\end{itemize}

To obtain the entropy expressions (\ref{bun2g9})-(\ref{bun2g11}) of
the above solution for $k\neq 0$, it is helpful to convert first to
$Y_{+}$
\be
S\!-\!S_{o}\!=\!\int \!r_{+}T^{-1}\!\cdot\!
\frac{1}{r_{+}}\frac{dM}{dY_{+}}dY_{+},
\ee
where we can find
\be
r_{+}T^{-1}\!=\!\frac{\pi e^{2}
\eta}{\xi}\frac{e^{6Y_{+}}(1\!+\!6Y_{+})}{6Y_{+}^{2}\!+\!5Y_{+}\!+\!1\!-\!\frac{\eta\sigma}{12\xi^{2}}}
\ee
\be
\frac{1}{r_{+}}\frac{dM}{dY_{+}}=\frac{k}{4e^{2}G}
sgn(\frac{k}{r_{+}^{2}}\!+\!\frac{\sigma}{6\xi})
\frac{(1\!+\!6Y_{+})(6Y_{+}^{2}\!+\!5Y_{+}\!+\!1\!-\!\frac{\eta\sigma}{12\xi^{2}})}{e^{6Y_{+}}
[Y_{+}(1\!+\!3Y_{+})\!-\!\frac{\eta\sigma}{24\xi^{2}}]^{2}},
\ee
so
finally,
\be
S-S_{o}=\frac{k\eta\pi}{4\xi G}
sgn\Big(\!\frac{k}{r_{+}^{2}}\!+\!\frac{\sigma}{6\xi}\!\Big)
\int\frac{(1\!+\!6Y_{+})^{2}}{[Y_{+}(1\!+\!3Y_{+})\!-\!\frac{\eta\sigma}{24\xi^{2}}]^{2}}dY_{+}.
\ee

\section{The general solution\label{gen}}

In this appendix we describe first the general solution of equation
(\ref{bun3a}) that we reproduce here for convenience.
\be
(3\zeta\!+\!8\eta)r^{2}g'^{\,2}+4(\zeta\!+\!4\eta)r g
g'+4(\zeta\!+\!2\eta)g^{2}+\frac{1}{2}\Big(\!4\sigma\!-\!\frac{3\xi^{2}}{\zeta\!+\!3\eta}\Big)r^{4}=0.
\label{abun3a1}
\ee

We change variables as
 \be
h(R)=e^{\frac{2(\zeta+4\eta)}{3\zeta+8\eta}R}g(R)=r^{\frac{2(\zeta+4\eta)}{3\zeta+8\eta}}g
\,\,\,\,,\,\,\,\,R=\ln{r}, \label{abun5}
\ee
 with dot being the derivative with respect to $R$.
We also introduce
\be
A=\frac{8\zeta(\zeta\!+\!3\eta)}{(3\zeta\!+\!8\eta)^{2}}\,\,,\,\,
B=\frac{1}{3\zeta\!+\!8\eta}
\Big(\!\frac{3\xi^{2}}{2(\zeta\!+\!3\eta)}\!-\!2\sigma\!\Big)\,\,,\,\,
C=\frac{16(\zeta\!+\!3\eta)}{3\zeta\!+\!8\eta}\sp A={1\over
4}C(6-C),\label{abun7}
\ee
so that equation (\ref{abun3a1}) is
\be
r^{2}g'^{\,2}+(C-4)rgg'+{1\over 2}(8-C)g^2-Br^4=0.
\ee

Under (\ref{abun5}) equation (\ref{abun3a1}) becomes
\be
\dot{h}^{2}+A h^{2}-B e^{CR}=0. \label{abun6}
\ee

From (\ref{abun6}) we observe  that
$Be^{CR}\!-\!Ah^{2}=Br^{C}\!-\!Ah^{2}\geq 0$. If $W(R)$ satisfies
the first order differential equation
\be
\dot{W}=\Big(\!1-\frac{C}{2}W\!\Big)(1+AW^{2}), \label{abun8}
\ee
then, $h(R)$ is given as
\be
 h(R)^{2}= \frac{Be^{CR}}{A\!+\!W^{-2}}\,\,.
\label{abun9}
\ee
This is the {\em general solution} for $h(R)$.
This is shown by defining
\be
W(R)\equiv {h(R)\over \dot{h}(R)}
\ee
and showing that equation (\ref{abun6}) implies that  $W(R)$
satisfies equation (\ref{abun8}), and $h(R)$ is given by
(\ref{abun9}).

 Since (\ref{abun9}) gives $h^{2}$, only
$|h|$ will be determined from the solution, and therefore, two
branches will always arise for each solution $|h|$. From
(\ref{abun9}), we must have $B/(A\!+\!W^{-2})\geq 0$.

 Equation (\ref{abun8}) leads to a quadrature
\be
\int\frac{dW}{(1\!-\!\frac{C}{2}W)(1\!+\!AW^{2})}=R-R_{o}=\ln\frac{r}{r_{o}},
\label{abun10}
\ee
where $r_{o}>0,R_{o}$ are integration constants.
The integrant can be broken as
 \be
\frac{1}{(1\!-\!\frac{C}{2}W)(1\!+\!AW^{2})}=\frac{1}{4A\!+\!C^{2}}\Big(\!\frac{C^{2}}{1\!-\!\frac{C}{2}W}
+\frac{2ACW}{1\!+\!AW^{2}}+\frac{4A}{1\!+\!AW^{2}}\!\Big).
\label{abun11}
\ee
 The integration of the first two terms in
(\ref{abun11}) gives
\be
\frac{1}{6}\ln{\frac{|1+AW^{2}|}{(1-\frac{C}{2}W)^{2}}}\,.
\ee

 The
integration of the third term depends on the sign of $A$. There are
two cases, one for $A>0$ and another for $A\leq 0$.

Before we proceed to consider the two cases, we should also consider
the second equation, namely (\ref{bun25}) that in terms of the
current variables, its coefficients translate as
\be
\hat{A}_{11}=(3\zeta\!+\!8\eta){rg\over W}\sp
\hat{B}_{11}=2(3\zeta\!+\!8\eta){Ag\over
CW}\Big(\!1\!-\!\frac{C}{2}W\!\Big)^{\!2}.
\ee
 Therefore,
equation (\ref{bun25}) is rewritten as
\be
\frac{d\ln\hat{N}}{dW}=-\frac{2A}{C}\,\frac{1-\frac{C}{2}W}{1+AW^{2}}\,\,.
\label{abun11a}
\ee

We now consider the two cases separately.

\subsection{$A>0$}

Equation (\ref{abun7}) implies that $0<C<6$ and a solution exists
only for $B>0$. The solution for $W$ is
\be
\ln\Big[\Big(\frac{r_{o}}{r}\Big)^{\!6}\frac{1+AW^{2}}{(1-\frac{C}{2}W)^{2}}\Big]+
\frac{4\sqrt{A}}{C}\arctan{(\!\sqrt{A}\,W\!)}=0, \label{abun12}
\ee
and for $h(r)$  we have the implicit equation
\be
\ln\Big[\Big(\frac{r}{r_{o}}\Big)^{\!3}
\,\Big|\!\sqrt{1\!-\!\frac{A}{B}\,\frac{h^{2}}{r^{C}}}-\e\frac{C}{2\sqrt{B}}\frac{h}{r^{C/2}}\Big|\Big]
=\frac{2\e\sqrt{A}}{C}\,\arcsin\!\Big(~\sqrt{\frac{A}{B}}\,\frac{h}{r^{C/2}}\!\Big).
\label{abun13}
\ee
with $h$ taking both positive and negative values
and $\e=\pm 1$.

Using  $\frac{2(\zeta+4\eta)}{3\zeta+8\eta}\!=\!\frac{C}{2}\!-\!2$,
we may rewrite the solution (\ref{abun13}) as an implicit equation
for  $g(r)$ as
\be
\Big(\frac{r}{r_{o}}\Big)^{\!3}
\,\Big|\!\sqrt{1\!-\!\frac{A}{B}\,\frac{g^{2}}{r^{4}}}-\e\frac{C}{2\sqrt{B}}\frac{g}{r^{2}}\Big|
=\exp\left[\frac{2\e\sqrt{A}}{C}\,\arcsin\!\Big(~\sqrt{\frac{A}{B}}\,\frac{g}{r^{2}}\!\Big)\right].
\label{abon14}
\ee

Integration of equation (\ref{abun11a}) gives
\be
\ln{\frac{\hat{N}(r)}{\hat{N}_{o}}}=\frac{1}{2}\ln(1\!+\!AW^{2})-\frac{2\sqrt{A}}{C}\arctan(\sqrt{A}\,W)\,,
\label{abun15}
\ee
and using (\ref{abun12}) we obtain
\be
\hat{N}(r)=\hat{N}_{o}\Big(\frac{r_{o}}{r}\Big)^{\!3}\,\frac{1+AW^{2}}{|1-\frac{C}{2}W|}\,,
\label{abun15a}
\ee
and  in terms of $g(r)$
\be
\hat{N}(r)=\Big(\frac{r_{o}}{r}\Big)^{\!3} \,\frac{\hat{N}_{o}}
{\sqrt{1\!-\!\frac{A}{B}\frac{g^{2}}{r^{4}}}\,\,
\Big|\!\sqrt{1\!-\!\frac{A}{B}\frac{g^{2}}{r^{4}}}-
\e\frac{C}{2\sqrt{B}}\frac{g}{r^{2}}\Big| }=\hat{N}_{o}
\frac{e^{-\frac{2\e\sqrt{A}}{C}\arcsin(\sqrt{\frac{A}{B}}\,\frac{g}{r^{2}})}}
{\sqrt{1\!-\!\frac{A}{B}\frac{g^{2}}{r^{4}}}} \,\,. \label{abon15b}
\ee
Finally, we choose for convenience the constant of integration
as
\be
\hat{N}_{o}=\sqrt{\frac{C}{6}}\,\exp\Big[\e
(sgng)\frac{2\sqrt{A}}{C}\arcsin\!\sqrt{\frac{2A}{3C}}\Big]\,,
\ee
so that the final solution becomes
\be
\hat{N}(r)=\sqrt{\frac{C}{6}}\,\Big(\frac{r_{o}}{r}\Big)^{\!3}\,
\frac{e^{\frac{2\e(sgng)\sqrt{A}}{C}\arcsin\!\sqrt{\frac{2A}{3C}}}}
{\sqrt{1\!-\!\frac{A}{B}\frac{g^{2}}{r^{4}}}\,\,
\Big|\!\sqrt{1\!-\!\frac{A}{B}\frac{g^{2}}{r^{4}}}-\e
\frac{C}{2\sqrt{B}}\frac{g}{r^{2}}\Big| } =\sqrt{\frac{C}{6}}\,\,
\frac{e^{-\frac{2\e\sqrt{A}}{C}\,[\,\arcsin(\sqrt{\frac{A}{B}}\,\frac{g}{r^{2}})-(sgng)\arcsin\!\sqrt{\frac{2A}{3C}}\,]}}
{\sqrt{1\!-\!\frac{A}{B}\frac{g^{2}}{r^{4}}}} \,\,. \label{abon15c}
\ee

 Equations (\ref{abon14}), (\ref{abon15c}) form the
general solution of the system for $A>0$.

For large distance $r\!>\!>\!r_{o}$ (which makes sense for $\e
(sgng)>0$), the leading behavior of (\ref{abon14}) for the solutions
with positive mass is
\be
{g\over r^2}\simeq
\e\sqrt{\frac{2B}{3C}}-\frac{\sqrt{B}\,r_{o}^{3}}{3r^{3}}\exp\!\Big(\frac{2\sqrt{A}}{C}\arcsin{\!\sqrt{\frac{2A}{3C}}}\Big)
+{\cal O}(r^{-6}).
\ee
We therefore obtain
\be
 f(r)\simeq
k+\frac{1}{4(\zeta\!+\!3\eta)}\Big[\xi\!+\tilde{\e}\sqrt{\xi^{2}\!-\!\frac{4\sigma}{3}(\zeta\!+\!3\eta)}\Big]\,r^{2}
-\frac{2GM}{ r}+{\cal O}(r^{-4})\,, \label{abun15c}
\ee
with
\be
{2GM}=\frac{\sqrt{B}\,r_{o}^{3}}{3}\exp\!\Big(\frac{2\sqrt{A}}{C}\arcsin{\!\sqrt{\frac{2A}{3C}}}\Big)\,\,\,\,\,\,
,\,\,\,\,\,\, \tilde{\e}=\e\, sgn(\zeta\!+\!3\eta).
\ee
Moreover we
have
\be
\hat{N}(r)=1+\mathcal{O}(r^{-6}).
\ee

To obtain the entropy expression (\ref{bun23}) for $k\neq 0$ it is
helpful to convert to $W_{+}$,
\be
S-S_{o}\!=\!\int
r_{+}T^{-1}\cdot\frac{1}{r_{+}}\frac{dM}{dW_{+}}dW_{+},
\ee
where we
can find
\be
r_{+}T^{-1}\!=\!4\pi\sqrt{\frac{6}{C}}\,\Big[\frac{\xi\sqrt{1\!+\!AW_{+}^{2}}}{2(\zeta+3\eta)}
+\e
\sqrt{\!B}\,(1\!-\!\frac{C\!-\!4}{2}W_{+})\Big]^{-1}\,e^{\frac{2\sqrt{A}}{C}
[\,\arctan(\sqrt{\!A}\,W_{+})-\arcsin\!\sqrt{2A/3C}\,]}\ee \be
\frac{1}{r_{+}}\frac{dM}{dW_{+}}=\frac{\sqrt{B}}{4G}\frac{k}{1\!+\!AW_{+}^{2}}
\Big[\!\sqrt{\!B}\,(1-\frac{C\!-\!4}{2}W_{+})+\e\frac{\xi\sqrt{1\!+\!AW_{+}^{2}}}{2(\zeta\!+\!3\eta)}
\Big]
\,\frac{e^{\frac{2\sqrt{A}}{C}[\arcsin\!\sqrt{2A/3C}-\arctan(\!\sqrt{\!A}\,W_{+})]}}
{\Big[\frac{\xi\sqrt{1+AW_{+}^{2}}}{4(\zeta+3\eta)}+\e
\sqrt{\!B}\,W_{+}\!\Big]^{2}},
\ee
so finally,
\be
S-S_{o}=\e
\frac{3\pi
k}{G}\sqrt{\frac{2B}{3C}}\int\frac{dW_{+}}{1\!+\!AW_{+}^{2}}
\Big[\frac{\xi\sqrt{1\!+\!AW_{+}^{2}}}{4(\zeta\!+\!3\eta)}+\e
\sqrt{\!B}\,W_{+}\!\Big]^{\!-2}\,.
\ee
This integration can be done
explicitly in terms of $W_{+}$ defining $\tan\theta=\sqrt{A}\,W_{+}$
and replacing in terms of $r_{+}$.

\subsection{$A\leq 0$}

In this case we have $C\geq 6$ or $C<0$. The solution from
(\ref{abun10}) is
\be
\frac{|1\!+\!AW^{2}|}{(1\!-\!\frac{C}{2}W)^{2}}\,\,
\Big|\!\frac{1\!-\!\sqrt{-A}\,W}{1\!+\!\sqrt{-A}\,W}\!\Big|^{\frac{2\sqrt{-A}}{C}}\!=\!
\Big(\frac{r}{\tilde{r}_{o}}\Big)^{6}, \label{abun16}
\ee
and in
terms of $h(r)$
\be
\Big|\!\sqrt{B\!-\!A\frac{h^{2}}{r^{C}}}-\e\frac{C}{2}\frac{h}{r^{C/2}}\!\Big|
\,\,\Big|\!\sqrt{B\!-\!A\frac{h^{2}}{r^{C}}}+\e\sqrt{-A}\frac{h}{r^{C/2}}\!\Big|^{\!\frac{2\sqrt{-A}}{C}}
\!=\!\Big(\frac{r_{o}}{r}\Big)^{3}, \label{abun17}
\ee
 where $\e=\pm 1$, $\tilde{r}_{o}>0$ is the integration constant, and
$r_{o}^{3}=\tilde{r}_{o}^{3}\,|B|^{\frac{1}{2}+\frac{\sqrt{-A}}{C}}$.
\newline
Finally, the solution (\ref{abun17}) is written in terms of $g(r)$
as
\be
\Big|\!\sqrt{B\!-\!A\frac{g^{2}}{r^{4}}}-\e\frac{C}{2}\frac{g}{r^{2}}\!\Big|
\,\,\Big|\!\sqrt{B\!-\!A\frac{g^{2}}{r^{4}}}+\e\sqrt{-A}\frac{g}{r^{2}}\Big|^{\!\frac{2\sqrt{-A}}{C}}
\!=\!\Big(\frac{{r}_{o}}{r}\Big)^{3},\label{abon18}
\ee
where $g$ is
allowed to take both positive and negative values.

Integration of equation (\ref{abun11a}) gives
\be
\frac{\hat{N}(r)}{\hat{N}_{o}}=\sqrt{|1+AW^{2}|}\,\,
\Big|\!\frac{1\!+\!\sqrt{-A}\,W}{1\!-\!\sqrt{-A}\,W}\!\Big|^{\frac{\sqrt{-A}}{C}}\,,
\label{abun18a}
\ee
 and using (\ref{abun16}) we obtain
  \be
\frac{\hat{N}(r)}{\hat{N}_{o}}=\Big(\frac{\tilde{r}_{o}}{r}\Big)^{\!3}\,\,\Big|\frac{1+AW^{2}}{1-\frac{C}{2}W}\Big|\,,
\label{abun18b}
\ee
which can be rewritten in terms  of $g(r)$ as
\be
\frac{\hat{N}(r)}{\hat{N}_{o}}=\Big(\frac{\tilde{r}_{o}}{r}\Big)^{\!3}
\,\,\frac{|B|}{\sqrt{B\!-\!A\,\frac{g^{2}}{r^{4}}}\,\,\Big|\!\sqrt{B\!-\!A\,\frac{g^{2}}{r^{4}}}-
\e\frac{C}{2}\frac{g}{r^{2}}\Big|}=
\frac{\Big|\!\sqrt{B\!-\!A\,\frac{g^{2}}{r^{4}}}+
\e\sqrt{-A}\,\frac{g}{r^{2}}\Big|^{\frac{2\sqrt{-A}}{C}}}{|B|^{\frac{\sqrt{-A}}{C}-\frac{1}{2}}\,\,
\sqrt{B\!-\!A\,\frac{g^{2}}{r^{4}}}}\,\,. \label{abon18c}
\ee

Choosing
$\hat{N}_{o}=\sqrt{\frac{|C|}{6}}\,\,(\sqrt{\frac{|C|}{6}}\,|1+\frac{2\sqrt{-A}}{C}|)^{-\frac{2\sqrt{-A}}{C}}$,
we obtain
\be
 \hat{N}(r)=\Big(\frac{r_{o}}{r}\Big)^{\!3}
\,\,\frac{\,\sqrt{\frac{|BC|}{6}}\,\,(\sqrt{\frac{|BC|}{6}}\,|1+\frac{2\sqrt{-A}}{C}|)^{-\frac{2\sqrt{-A}}{C}}}
{\sqrt{B\!-\!A\,\frac{g^{2}}{r^{4}}}\,\,\Big|\!\sqrt{B\!-\!A\,\frac{g^{2}}{r^{4}}}-
\e\frac{C}{2}\frac{g}{r^{2}}\Big|}=
\frac{\sqrt{\frac{|BC|}{6}}\,\,\Big|\!\sqrt{B\!-\!A\,\frac{g^{2}}{r^{4}}}+\e
\sqrt{-A}\,\frac{|g|}{r^{2}}\Big|^{\frac{2\sqrt{-A}}{C}}}{(\sqrt{\frac{|BC|}{6}}\,
|1\!+\!\frac{2\sqrt{-A}}{C}|)^{\frac{2\sqrt{-A}}{C}}\,\,
\sqrt{B\!-\!A\,\frac{g^{2}}{r^{4}}}}\,\,. \label{abon18d}
\ee

 Equations (\ref{abon18}), (\ref{abon18d}) form the general
solution of the system for $A\leq 0$.
\bigskip

For large distance $r\!>\!>\!r_{o}$ (for $C\geq 6$, $B>0$ or $C<0$,
$B<0$) the leading behavior of (\ref{abon18}) for the solutions with
positive mass is
\be
g(r)\simeq \e\sqrt{2B\over 3C}~r^2-{2GM\over
r}+{\cal O}(r^{-4})\sp 2GM= \frac{r_o^3}{ 3}\Big[\sqrt{BC\over
6}\,\Big|1\!+\!{2\sqrt{|A|}\over C}\Big|\Big]^{-\frac{2\sqrt{|A|}}{
C}}.
 \label{t111}
\ee
We therefore obtain
\be
 f(r)\simeq
k+\frac{1}{4(\zeta\!+\!3\eta)}\Big[\xi\!+\tilde{\e}\sqrt{\xi^{2}\!-\!\frac{4\sigma}{3}(\zeta\!+\!3\eta)}\Big]\,r^{2}
-\frac{2GM}{ r}+{\cal O}(r^{-4})\,\,\,\,\,\, ,\,\,\,\,\,\,
\tilde{\e}=\e\, sgn(\zeta\!+\!3\eta). \label{abun15cc}
\ee
Moreover
we have
\be
\hat{N}(r)=1+\mathcal{O}(r^{-6}).
\ee

To obtain the entropy expression (\ref{bun244}) for $k\neq 0$ it is
helpful to convert to $W_{+}$,
\be
S-S_{o}\!=\!\int
r_{+}T^{-1}\cdot\frac{1}{r_{+}}\frac{dM}{dW_{+}}dW_{+},
\ee
where we
can find
\be
r_{+}T^{-1}\!=\!\frac{4\pi}{\hat{N}_{o}}\Big[\frac{\xi\sqrt{1\!+\!AW_{+}^{2}}}{2(\zeta+3\eta)}+\e
\sqrt{B}\,\Big(\!1\!-\!\frac{C\!-\!4}{2}W_{+}\!\Big)\Big]^{\!-1}\Big(\frac{1\!-\!\sqrt{-A}\,W_{+}}{1\!+\!\sqrt{-A}\,W_{+}}\Big)
^{\!\frac{\sqrt{-A}}{C}}
\ee
\be
\frac{1}{r_{+}}\frac{dM}{dW_{+}}=\frac{\hat{N}_{o}}{2G}\sqrt{\frac{3B}{2C}}\,\frac{k}{1\!+\!AW_{+}^{2}}
\,\Big[\sqrt{B}\,\Big(\!1\!-\!\frac{C\!-\!4}{2}W_{+}\!\Big)+\e
\frac{\xi\sqrt{1\!+\!AW_{+}^{2}}}{2(\zeta+3\eta)}\Big]\,
\frac{(\frac{1+\sqrt{-A}\,W_{+}}{1-\sqrt{-A}\,W_{+}})^{\frac{\sqrt{-A}}{C}}}
{\Big[\!\frac{\xi\sqrt{1+AW_{+}^{2}}}{4(\zeta+3\eta)}+\e
\sqrt{B}\,W_{+}\!\Big]^{2}}\,\,,
\ee
so finally,
\be
S-S_{o}=\e
\frac{3\pi k}{G}\sqrt{\frac{2B}{3C}}\int\!
\frac{dW_{+}}{1\!+\!AW_{+}^{2}}\Big[\frac{\xi\sqrt{1\!+\!AW_{+}^{2}}}{4(\zeta+3\eta)}+\e
\sqrt{B}\,W_{+}\Big]^{-2}\,.
\ee
This integration can be done
explicitly in terms of $W_{+}$ defining
$u=\tan(\theta/2),\,\sin\theta=\sqrt{-A}\,W_{+}$ and replacing in
terms of $r_{+}$.

\section{Dispersive geodesics\label{geo}}

Particles with non-standard dispersion relations, and in particular
non-Lorentz invariant ones, do not follow the usual geodesics in
the gravitational field. This has been discussed in several very recent works \cite{poly}, \cite{suyama}, \cite{rama}.
We will follow a slightly different approach here to study dispersive geodesics.
We will eventually investigate the fate of the horizon for such particles
and find that it seems to disappear, as intuition suggests.

To proceed, we write the   metric in the ADM form
\be
G_{00}=-N^2+N_{i}g^{ij}N_j\sp G_{0i}=N_i\sp G_{ij}=g_{ij}\sp \det[G]=\det[g] N^2
\label{p1}
\ee
as well as the inverse metric
\be
G^{00}=-{1\over N^2}\sp G^{0i}={N^i\over N^2}\sp G^{ij}=g^{ij}-{N^iN^j\over N^2}
\label{p2}
\ee
where indices of $N_i,g_{ij}$ are raised with the spatial metric $g_{ij}$.

We will also consider a scalar mode which in the standard relativistic case has a coupling to gravity in ADM notation given by
\be
S=\int d^3xdt \sqrt{g}N\left[-{1\over N^2}(\dot \Phi-N^i\partial_i\Phi)^2+g^{ij}\partial_i\Phi \partial_j\Phi+V(\Phi)\right]
\label{p3}
\ee
For the non-relativistic case we write a  general non-relativistic quadratic action in the Lifshitz spirit as
\be
S_{\rm nr}=\int d^3xdt \sqrt{g}N\left[-{1\over N^2}(\dot \Phi-N^i\partial_i\Phi)^2-\Phi F[\square]\Phi\right]\sp \square =g^{ij}\nabla_i \nabla_j
\label{p4}
\ee
with equations of motion
\be
\hat H~\Phi={1\over N\sqrt{g}}\partial_t\left[{\sqrt{g}\over N}(\dot \Phi-N^i\partial_i\Phi)\right]-{1\over N\sqrt{g}}\partial_j\left[\sqrt{g}{N^j\over N}(\dot \Phi-N^i\partial_i\Phi)\right]
-F[\square] \Phi=0
\label{p5}
\ee
where in the spatial part we neglected derivatives acting on $g_{ij}, N$ as they will be
irrelevant in the derivation of the geodesic (geometric optics) equations.

To pass to point-like trajectories, we replace $i\pa_t\to p_0$, $i\pa_i\to p_i$, and neglect metric derivatives to obtain
the equivalent Hamiltonian (``zero energy") constraint
\be
H=-{(p_0-N^ip_i)^2\over N^2}+F[\zeta]=0\sp \zeta=g^{ij}p_ip_j
\label{e}
\ee
This is implemented with a Lagrange multiplier $e$, to obtain the world-line action as
\be
S_{wl}=\int_0^1d\tau\left[p_0 \dot t+p_i\dot x^i+{e\over 2} H\right]
\label{p9}
\ee
where $\tau$ is the affine time of the path and dot stands for $\pa_{\tau}$.
$e$ is a one-dimensional einbein.
The action (\ref{p9}) is invariant under general reparametrization of the affine time
\be
\tau'=f(\tau)\sp e'={e\over f'(\tau)}
\ee
Using such reparametrizations the einbein can be set to a constant $T$.

The classical path equations come from varying $e$, $t,x^i$, and $p_0,p_i$.
The variation of $e$ by definition gives (\ref{e}) while the momentum variations give
\be
\dot t+{e\over 2}{\delta H\over \delta p_0}=\dot t-{e\over N^2}(p_0-N^ip_i)=0~~~\to~~~ p_0-N^ip_i={N^2\over e}\dot t
\label{p0}
\ee
\be
\dot x^i+{e\over 2}{\delta H\over \delta p_i}=\dot x^i+{eN^i\over N^2}(p_0-N^ip_i)+
eF'(\zeta)g^{ij}p_j=0~~~\to~~~\dot x^i+N^i\dot t=-eF'(\zeta)~g^{ij}p_j
\label{pi}
\ee
Finally the variations over the coordinates give
\be
\dot p_0={e\over 2}{\pa H\over \pa t}\sp \dot p_i={e\over 2}{\pa H\over \pa x^i}
\label{x}
\ee
Equation (\ref{pi}) gives
\be
\zeta (F'(\zeta))^2={g_{ij}(\dot x^i+N^i\dot t)(\dot x^j+N^j\dot t)\over e^2}\equiv {\xi\over e^2}
\label{191}
\ee
This is a non-linear equation in general, and  we can pick a  solution $\zeta(\xi/e^2)$.
Then we can express
\be
p_i=-{g_{ij}(\dot x^j+N^j\dot t)\over e F'(\zeta)}\sp p_0={N^2\over e}\dot t-{N_i(\dot x^i+N^i \dot t)\over e F'(\zeta)}
\label{p10}
\ee

We also obtain by substitution
\be
p_0 \dot t+p_i\dot x^i={N^2\over e}\dot t^2-{g_{ij}(\dot x^i+N^i\dot t)(\dot x^j+N^j\dot t)\over eF'(\zeta)}
={N^2\over e}\dot t^2-e\zeta F'(\zeta)
\label{p11}
\ee
where in the last equation we used (\ref{191}).
We also have
\be
H=-{N^2\over e^2}\dot t^2+F(\zeta)
\label{p12}
\ee
and finally the on-shell action is
\be
S_{wl}={1\over 2}\int_0^1d\tau\left[{N^2\over e}\dot t^2+e(F(\zeta)-2\zeta F'(\zeta))\right]
\label{p13}
\ee
where $\zeta$ should be thought as a function of the metric, $\dot x^i$, and the einbein as given from (\ref{191}).
Indeed varying $e,x^i,t$ using the constraint (\ref{1})gives that appropriate geodesic equations and the zero energy constraint.
This action reduces to the standard relativistic action for $F(\zeta)=\zeta+m^2$.

We now consider  a particle with a non-standard  dispersion relation $F(\zeta)=\zeta^n$ (corresponds to a dispersion relation
 $(p^0)^2-(\vec p^{\,2})^n=0$) a radial geodesic and a spherical symmetric BH metric
to obtain
\be
S={1\over 2}\int_0^1 d\tau\left[{N^2\dot t^2\over e}+(1-2n)e^{-{1\over 2n-1}}\left({\dot r\over 2n\sqrt{f}}\right)^{2n\over 2n-1}\right]
\label{p27}
\ee
\be
e=\left({2N\over E}\right)^{2n-1\over n}{\dot r\over 2n\sqrt{f}}\sp \dot t={2\over E}\left({2N\over E}\right)^{-{1\over n}}{\dot r\over 2n\sqrt{f}}
\label{p28}
\ee
where $E$ is a constant of integration (energy).

From (\ref{p2}) we also obtain

\be
{dr\over dt}=nE\sqrt{f}\left({2N\over E}\right)^{1\over n}
\label{p30}
\ee
For any $n>1$ the geodesic is regular at the zeros of $f,N$.

If at $r_*$ $f$ has an a-th order zero and $N^2$ a b-th order zero, then the geodesic is singular if
\be
a+{b\over n}\geq 2.
\label{p31}
\ee
We have always $b\geq a$. In the case of HL we expect generically that $n=3$, therefore the condition is $a+b/3\geq 2$.
For a single zero of $f$, $a=1$ and then $b\geq 3$ that is unlikely. Generically $b=a=1$.

For a double zero of $f$, then we always have a singularity.
This suggests that  only extremal BH have a horizon in support of the intuitive notion that if a particle has a
 dispersion relation with an infinite speed of propagation,
then it feels no horizon in the background of BH solutions.

\end{document}